\documentclass[12pt]{article}
\usepackage{amsfonts,amsmath, amssymb,epsfig}
\usepackage[latin1]{inputenc}
\usepackage[english]{babel}
\usepackage{hyperref}
\usepackage{caption}
\usepackage{subcaption}
\usepackage{color}
\newtheorem{thm}{Th\'eor\`eme}[section]
\newtheorem{cor}[thm]{Corollaire}
\newtheorem{lem}[thm]{Lemme}
\newtheorem{pro}[thm]{Proposition}
\newtheorem{dfn}[thm]{D\'efinition}
\newtheorem{rmk}[thm]{Remark}
\newtheorem{expl}[thm]{Exemple}

\def\dessous#1\sous#2{\mathrel{\mathop{\kern0pt#2}\limits_{#1}}}

\newcommand{\R}{\mathbb R}
\newcommand{\C}{\mathbb C}
\newcommand{\N}{\mathbb N}

\newcommand{\1}{1 \! \! {\rm I}}

\newcommand{\beq}{\begin{eqnarray}}
\newcommand{\eeq}{\end{eqnarray}}
\newcommand{\bpro}{\begin{pro}}
\newcommand{\epro}{\end{pro}}
\newcommand{\blem}{\begin{lem}}
\newcommand{\elem}{\end{lem}}
\newcommand{\bdfn}{\begin{dfn}}
\newcommand{\edfn}{\end{dfn}}
\newcommand{\bcor}{\begin{cor}}
\newcommand{\ecor}{\end{cor}}
\newcommand{\bthm}{\begin{thm}}
\newcommand{\ethm}{\end{thm}}
\newcommand{\bex}{\begin{expl}}
\newcommand{\eex}{\end{expl}}
\newcommand{\brmk}{\begin{rmk}}
\newcommand{\ermk}{\end{rmk}}
\newcommand{\benum}{\begin{enumerate}}
\newcommand{\eenum}{\end{enumerate}}
\newcommand{\bitem}{\begin{itemize}}
\newcommand{\eitem}{\end{itemize}}
\def\deq{\stackrel{\mathrm{def}}{=}}

\begin{document}
\begin{center}
{\Large\bf {Coherent states  for the  exotic  Landau problem and related properties }}

 \vspace{0.5cm}

 Isiaka Aremua

 {\em
Laboratoire de Physique des Mat\'{e}riaux et des Composants \`{a} Semi-Conducteurs, Facult\'{e} Des Sciences (FDS),
D\'{e}partement de Physique, Universit\'{e} de Lom\'{e} (UL), 01 B.P. 1515 Lom\'{e} 01, Togo}\\
{\em University of Abomey-Calavi, International Chair in Mathematical Physics \\
			and Applications (ICMPA), 072 B.P. 050 Cotonou, Benin} \\
E-mail: claudisak@gmail.com

\vspace{1.0cm}

\today

\begin{abstract}
\noindent
This work presents a comprehensive study of the exotic Landau model in a two-dimensional noncommutative plane. Beginning with the classical formulation where two conserved quantities $\mathcal{P}_i$ and $\mathcal{K}_i$ are derived, we proceed to the quantum level where these lead to two independent oscillator representations generating bosonic Fock spaces $\Gamma_{\mathcal{P}}$ and $\Gamma_{\mathcal{K}}$. Coherent states satisfying all Klauder's criteria are explicitly constructed, and their physical properties including normalization, continuity, resolution of the identity, temporal stability, and action identity are rigorously proven. We further develop matrix vector coherent states and quaternionic vector coherent states, examining their mathematical structure and physical implications. Detailed calculations of the free particle propagator via path integrals, uncertainty relations, and time evolution of probability densities are provided.
\end{abstract}

\end{center}


\setcounter{footnote}{0}
\section{Introduction}
In quantum field theory  literature,   the natural appearance of noncommutativity in string theories has increasingly led to attempts to
study physical problems in
noncommutative spaces \cite{doplicher, douglas}.
Similar structures also arise in specific
approaches towards a theory of quantum gravity, such as M-theory in the presence of
background fields \cite{seiberg-witten} or tentative formulations of relativistic quantum theories of gravity
through spacetime noncommutativity \cite{conf}.
The description
of such   systems \cite{girvin} is adequately provided by the well known Landau model \cite{landau}.
See also \cite{dodonov} which makes an excellent review on quantum Hamiltonians related to this quantum model, and references listed therein. Since this discovery, the quantum states of a particle in a magnetic and  electromagnetic fields   on noncommutative plane
 \cite{horvathy}-\cite{iaremuaetal} (and also references quoted therein) and their quantum Hall limit \cite{girvin}   have been  attracting
considerable attention.
The  standard approach of the noncommutative Landau problem  consists
in  considering the commutation relations and Hamiltonian \cite{horvathy}
\beq
[x_{1}, x_{2}] &=& i \theta, \quad [x_{i}, p_{j}] = i \delta_{ij}, \quad [p_{1}, p_{2}] = i B, \quad
H = \frac{{\bf p}^{2}}{2M} + V
\eeq
with   $\theta$, $B$, ${\bf p}$, $V$ and $M$ being the noncommutative parameter, the  magnetic field, the momentum,
the electric potential and the mass,
respectively. Such a model has been studied
  in some previous works \cite{horvatetal}.
There is also  an \textquotedblleft exotic\textquotedblright version \cite{horvathy} of the same model, where the parameter
$M^* = M(1-B\theta)$, which plays the role of the effective mass, is
considered. Moreover,
 some similarities between the standard noncommutative approach
and the Peierls one  \cite{horvathy}  have been highlighted in the analysis of the noncommutative Landau problem.

Hilbert spaces are the skeleton of the mathematical structure for quantum theories. Within
this framework, coherent states (CSs)  represent a specific, overcomplete family of vectors that
offer a powerful bridge between quantum and classical descriptions.
For various generalizations, approaches, and their properties, one may consult
\cite{klauder-skagerstam,  perelomov, ali-antoine-gazeau} and references therein.
In the specific context of noncommutative geometry, CSs have proven to be
exceptionally useful tools \cite{grosse-presnajder}. Leveraging techniques developed for noncommutative quantum mechanics \cite{gouba-scholtz}, Gazeau-Klauder CSs \cite{gazeauklauder}  have been successfully constructed \cite{ben-scholtz}. In \cite{aremua-gouba1}, CSs for a system of an electron moving in a plane under uniform external magnetic and electric
fields, that fulfill  Gazeau-Klauder criteria, first in the context of discrete and continuous spectra and next by considering both spectra purely discrete, have been constructed.
Furthermore, for an electron in a uniform electromagnetic field coupled to a harmonic potential on the noncommutative plane, more elaborate structures such as matrix vector coherent
states (MVCSs) and quaternionic vector coherent states (QVCSs) have been built and analyzed \cite{a-hk}.
Besides, the density operator representation for Barut-Girardello CSs,  multi-matrix VCSs,  and also  two-component VCSs basis representation for a supersymmetric
harmonic oscillator,  have  been performed and applied to
Landau levels with their relevant mathematical  and statistical properties  derived and discussed \cite{aremuadensity}.
Recent work has explored the classical
exotic Landau problem with its two conserved quantities, leading to the construction of
entangled coherent states analogous to Bell states, with applications studied in quantum
information protocols like qubit teleportation \cite{aremua-gouba-qubit}.

This work presents a comprehensive study of the exotic Landau problem on the noncommutative plane, as defined in \cite{horvathy, horvatetal}. Our primary methodology is based on the
formalism developed in \cite{gouba-scholtz, a-hk}. We systematically construct CSs for this
model, ensuring they satisfy Klauder's minimal criteria. This coherent state  framework is
then employed to extract novel physical insights and perform non-trivial calculations. A key result derived from the completeness relation of these CSs is  the free particle propagator
within this noncommutative setting by
utilizing techniques adapted from  \cite{gouba-scholtz, gangopadhyay-scholtz}.
Our analysis reveals an ultraviolet cutoff intrinsically
induced by the noncommutative parameter $\theta$, a feature observed in studies of a
free particle on a noncommutative plane  \cite{hstan}. Beyond standard CSs, we further develop the formalism by constructing VCSs and QVCSs for this quantum model. We
investigate their mathematical structures and physical properties in detail. Additionally,
leveraging tools such as the Wigner transform and specific unitary mappings, we demonstrate
how an analogue of the VCSs built on the Hilbert space $\C^2 \otimes\mathcal{H}_q \otimes\mathcal{H}_q $ can be realized on the
space $\C^2 \otimes \mathfrak{H}^{\otimes 4} $,  where $\mathfrak{H} = L^2(\R)$. This mapping reveals deeper structural aspects of the
model's representation theory.

The paper is organized as follows. Section 2 introduces the physical model, including a chiral
decomposition of its Hamiltonian, a discussion of its eigenspectrum, and a description of the
pertinent quantum Hilbert space. Section 3 details the construction of CSs,
verifies the fulfillment of Klauder's criteria, and examines important physical implications
derived from this construction. Section 4 is devoted to the construction and analysis of
VCSs and QVCSs, including an investigation of their mathematical and physical properties.
The temporal evolution of the QVCSs is studied in Section 5. In Section 6, we explore the role of the Wigner transform and implement unitary mappings to construct related classes
of VCSs. Finally, there follow concluding remarks.
\section{The exotic Landau problem}\label{sec2}

This section provides a comprehensive derivation of the exotic Landau problem by integrating findings from previous studies \cite{horvathy, horvatetal, iaremuaetal} and including detailed intermediate steps for  clarity.
\subsection{The Model at the Classical Level}\label{ssec:classical}
We examine a two-dimensional noncommutative plane characterized by fundamental commutation relations given by
	\begin{equation}\label{fcr1}
		\{x_i, x_j \} = \theta \varepsilon^{ij}; \quad
		\{x_i, p_j\} = \delta^{ij}; \quad
		\{ p_i, p_j\} = 0,
	\end{equation}
	where \(\varepsilon^{ij}\) denotes the normalized antisymmetric tensor (\(\varepsilon^{12} = 1\), \(\varepsilon^{21} = -1\)), \(\delta_{ij}\) represents the Kronecker delta, and \(\theta\) is the noncommutative parameter. In this context, the associated Poisson bracket in phase space is modified from the canonical case by incorporating an additional term:
  \begin{equation}\label{pb_modified}
		\{f, g \} = \frac{\partial f}{\partial \vec{x}} \cdot \frac{\partial g}{\partial \vec{p}}
		- \frac{\partial g}{\partial \vec{x}} \cdot \frac{\partial f}{\partial \vec{p}}
		+ \theta \left( \frac{\partial f}{\partial x_1}\frac{\partial g}{\partial x_2}
		- \frac{\partial g}{\partial x_1}\frac{\partial f}{\partial x_2} \right).
	\end{equation}
	For a system comprising a charged particle with mass \( M \) and charge \( e \) moving in this plane, the noncommutative parameter \( \theta \) is regarded as exotic and is related to the exotic parameter \( \kappa \) by the following relation:
	\begin{equation}\label{theta_kappa}
		\theta = \frac{\kappa}{M^2}.
	\end{equation}
	The dynamics of the system are governed by the standard Hamiltonian:
	\begin{equation}\label{sol011}
		\mathcal{H} = \frac{1}{2M}\sum_{i=1}^{2} p_i^2 + e V(x_1, x_2), \quad i = 1, 2,
	\end{equation}
	where \( V \) represents the electric potential, which is assumed to be time-independent.

		In the presence of constant electromagnetic fields (\(\vec{E}\) and \(\vec{B}\)), the Hamiltonian in equation \eqref{sol011} remains unchanged; however, the Poisson bracket is modified to:
	\begin{equation}\label{pb_emfield}
		\{f, g \} = \frac{\partial f}{\partial \vec{x}} \cdot \frac{\partial g}{\partial \vec{p}}
		- \frac{\partial g}{\partial \vec{x}} \cdot \frac{\partial f}{\partial \vec{p}}
		+ \theta \left( \frac{\partial f}{\partial x_1}\frac{\partial g}{\partial x_2}
		- \frac{\partial g}{\partial x_1}\frac{\partial f}{\partial x_2} \right)
		+ B \left( \frac{\partial f}{\partial p_1}\frac{\partial g}{\partial p_2}
		- \frac{\partial g}{\partial p_1}\frac{\partial f}{\partial p_2} \right).
	\end{equation}
The fundamental commutation relations~\eqref{fcr1} become
	\begin{equation}\label{pbrack}
		\{x_i, x_j\} = \frac{M}{M^*}\theta \varepsilon^{ij}, \qquad
		\{x_i, p_j\} = \frac{M}{M^*}\delta^{ij}, \qquad
		\{p_i, p_j\} = \frac{M}{M^*}eB\varepsilon^{ij},
	\end{equation}
	where $\theta$ and the charge $e$ combine with the magnetic field $B$ to yield an effective mass
 $M^* = M(1 - e\theta B)$.
	We choose the vector potential as $A_i = \frac{1}{2}B\epsilon_{ij}x_j$ and the electric field as $E_i = -\partial_i V$. The equations of motion are derived from $\dot{\chi} = \{\mathcal{H}, \chi\}$, where $\chi \in \{x_1, x_2, p_1, p_2\}$ and $i = 1, 2$. Using the Poisson bracket~\eqref{pb_emfield} with $\mathcal{H} = \frac{p_i^2}{2M} + eV(\vec{x})$, we obtain
	\begin{align}
		\dot{x}_i &= \{x_i, \mathcal{H}\} = \frac{\partial x_i}{\partial \vec{x}} \cdot \frac{\partial \mathcal{H}}{\partial \vec{p}}
		- \frac{\partial \mathcal{H}}{\partial \vec{x}} \cdot \frac{\partial x_i}{\partial \vec{p}}
		+ \theta \left( \frac{\partial x_i}{\partial x_1}\frac{\partial \mathcal{H}}{\partial x_2}
		- \frac{\partial \mathcal{H}}{\partial x_1}\frac{\partial x_i}{\partial x_2} \right) \nonumber \\
		&\quad + B \left( \frac{\partial x_i}{\partial p_1}\frac{\partial \mathcal{H}}{\partial p_2}
		- \frac{\partial \mathcal{H}}{\partial p_1}\frac{\partial x_i}{\partial p_2} \right). \label{xdot_full}
	\end{align}
Using the identities \(\frac{\partial x_i}{\partial x_j} = \delta_{ij}\) and \(\frac{\partial x_i}{\partial p_j} = 0\), the first and last terms simplify. For the term involving \(\theta\), we note that \(\frac{\partial \mathcal{H}}{\partial x_j} = e \frac{\partial V}{\partial x_j} = -eE^j\). With \(\varepsilon^{12} = 1\) and \(\varepsilon^{21} = -1\), we arrive at the resulting expression after multiplying by \( M \) and rearranging:
	\begin{equation}\label{momentum-relation}
		p_i = M\dot{x}_i + Me\theta\varepsilon^{ij}E^j.
	\end{equation}
Similarly, for $\dot{p}_i = \{p_i, \mathcal{H}\} $ from the relations $\frac{\partial p_i}{\partial x_j} = 0$ and $\frac{\partial p_i}{\partial p_j} = \delta_{ij}$, and using  equation (\ref{momentum-relation}), we get,
	\begin{equation}\label{equa000}
		M^*\dot{x}_i = p_i - Me\theta\varepsilon^{ij}E^j, \qquad
		\dot{p}_i = eB\varepsilon^{ij}\dot{x}_j + eE^i, \quad i, j = 1, 2.
	\end{equation}
In the case of a purely magnetic field, differentiating the first equation of (\ref{equa000}) with \( E = 0 \) results in
\begin{equation}
		M^*\ddot{x}_i = \dot{p}_i = eB\varepsilon^{ij}\dot{x}_j \Longrightarrow \ddot{x}_i = \omega^*\varepsilon^{ij}\dot{x}_j
	\end{equation}
implying that the particle undergoes a modified cyclotronic motion with a frequency
$\omega^* = \frac{\omega}{1 -e\theta B}$, described by:
\begin{equation}
x_i(t) = R(- \omega^* t) \alpha_i + \beta_i
\end{equation}
where $\vec{\alpha} = (\alpha_1,\alpha_2)$ and $\vec{\beta} = (\beta_1,\beta_2)$ are constant vectors. The time-dependent translation (boost):
	\begin{equation}\label{boost_trans}
		x_i \to x_i + b_i, \quad p_i \to p_i + M^*\dot{b}_i
	\end{equation}
	is a symmetry of equation~\eqref{equa000} (with $\vec{E} = 0$) if and only if $\vec{b} = (b_1, b_2)$ satisfies:
	\begin{equation}\label{equa01}
		M^*\ddot{b}_i - eB\varepsilon^{ij}\dot{b}_j = 0 \Longrightarrow \ddot{b}_i = \omega^*\varepsilon^{ij}\dot{b}_j,
	\end{equation}
which has the general solution:
	\begin{equation}\label{boost_sol}
		b_i(t) = R(-\omega^* t)a_i + c_i,
	\end{equation}
	where $\vec{a} = (a_1, a_2)$ and $\vec{c} = (c_1, c_2)$ are constant vectors. The conserved quantities associated with these symmetries are
	\begin{equation}\label{conserved_P}
		\mathcal{P}_i = M^*(\dot{x}_i - \omega^*\varepsilon^{ij}x_j), \quad
		\mathcal{K}_i = \frac{M^*}{M}R(\omega^* t)p_i = \frac{{M^*}^2}{M}R(\omega^* t)\dot{x}_i, \quad i = 1, 2,
	\end{equation}
	where $R(\omega^* t)$ denotes the rotation by angle $\omega^* t$. These conserved quantities satisfy the following Poisson bracket algebra:
	\begin{equation}\label{algebra}
		\{\mathcal{P}_i, \mathcal{P}_j\} = -M^*\omega^*\varepsilon^{ij}, \quad
		\{\mathcal{K}_i, \mathcal{K}_j\} = (1-e\theta B)M^*\omega^*\varepsilon^{ij}, \quad
		\{\mathcal{P}_i, \mathcal{K}_j\} = 0.
	\end{equation}
\subsection{Model at the quantum level}
At the quantum level, classical quantities are promoted to operators, indicated by "hats," and Poisson brackets are replaced by commutators multiplied by the factor \(i\hbar\). Due to the exotic noncommutative parameter, the conventional position representation is not applicable here.

Under the condition $E = 0$ and $e B\theta \neq 1$, the quantum Hamiltonian
\begin{equation}\label{jjjj}
\hat H = \sum_{i = 1}^2\frac{\hat {p_i}^2}{2M},\quad i = 1,2,
\end{equation}
depends exclusively on the conserved quantities $\hat{\mathcal{K}}_i,  \: i = 1, 2$,  which satisfy the commutation relations:
\begin{equation}
[\hat{\mathcal{K}}_i,\;\hat{\mathcal{K}}_j]  = i\hbar(1 -e\theta B) M^* \omega^* \varepsilon^{ij}.
\end{equation}
The annihilation and creation operators $a, a^\dagger$ are defined as follows:
\begin{equation}\label{ladder00}
\hat a =    \hat{\mathcal K}^{1} + i\hat{\mathcal K}^{2}, \qquad  \hat a^{\dag} =
\hat{\mathcal K}^{1} - i   \hat{\mathcal K}^{2}, \quad[\hat a, \hat a^{\dag}] = 2\hbar (1-eB\theta) M \omega.
\end{equation}
The quantum hamiltonian  becomes (\ref{jjjj}) becomes
\begin{equation}\label{operat01}
\hat{H} = \frac{1}{2M(1-eB\theta)^{2}}\hat a^{\dag}\hat a + \frac{ \hbar \omega^{*}}{2}\,,
\end{equation}
where   $\omega^{*} = {eB}/{M^{*}},  M^{*} = (1-eB\theta)M$. It is convenient to introduce normalized operators $\{\mathfrak a,  {\mathfrak a}^{\dag}\}$  as follows

  \begin{equation}\label{algeb00}
 \mathfrak a
 = \frac{1}{ \sqrt{2\hbar (1-eB\theta)M\omega}}\, \hat a \quad
{\mathfrak a}^{\dag} =  \frac{1}{\sqrt{2\hbar (1-eB\theta)M\omega}} \hat a^{\dag}
\end{equation}
that satisfy the Fock algebra $
 [\mathfrak a,  {\mathfrak a}^{\dag}] = \mathbb {I}$. The noncommutative configuration space in this sector is thus isomorphic to the bosonic Fock space:

 \begin{equation}
 \Gamma_{\mathcal K} = \textrm{span} \left\lbrace
 | n\rangle \equiv \frac{1}{\sqrt{n!}}({\mathfrak{a}^\dagger})^n | 0 \rangle_{\mathcal K}\right\rbrace_{n= 0}^\infty.
 \end{equation}

Let's consider now the oscillator representation of the other conserved quantity, $\hat{\mathcal{P}}_i, \: i = 1,2$,
which are  \textquotedblleft ${\hat{x}_i, i = 1,2 }$-only operators \textquotedblright, as follows
\begin{equation}
\hat b = \hat{\mathcal P}^{1} + i \hat{\mathcal P}^{2}, \quad
\hat b^{\dag} = \hat{\mathcal P}^{1} - i \hat{\mathcal P}^{2}, \qquad  [\hat b, \hat b^{\dag}] = 2\hbar M\omega\,.
\end{equation}

In the same manner as above, it is convenient to introduce the operators
$\{\mathfrak b, \mathfrak b^{\dag}\}$
\begin{equation}
 \mathfrak b = \frac{1}{\sqrt{2\hbar M \omega}}\, \hat b, \quad
\mathfrak b^{\dag} =  \frac{1}{\sqrt{2\hbar M \omega }}\hat b^{\dag},
\end{equation}
that satisfy the Fock algebra $ [\mathfrak b, \mathfrak b^{\dag}] = \mathbb I $. The non-commutative configuration in this sector is then isomorphic to the boson Fock space
\begin{equation}
 \Gamma_{\mathcal P} = \textrm{span} \left\lbrace
 | m\rangle \equiv \frac{1}{\sqrt{n!}}({\mathfrak{b}^\dagger})^m | 0 \rangle_{\mathcal P}\right\rbrace_{m= 0}^\infty.
 \end{equation}

The total Fock space of the system is the tensor product
$\Gamma = \Gamma_{\mathcal P} \otimes \Gamma_{\mathcal K}$ whose basis is given by the coupled states:
\begin{equation}\label{fock000}
\Gamma = \textrm{span}\left\lbrace |m \rangle \otimes |n \rangle = |m, n\rangle \equiv \frac{1}{\sqrt{m !n !}}(\mathfrak {b}^{\dag})^{m}(\mathfrak {a}^{\dag})^{n}|0,0\rangle_{\mathcal{K}, \mathcal{P}}
\right\rbrace^{\infty}_{m,n =0}\,.
\end{equation}

The system's energy depends solely on the dynamics linked to $\mathcal{K}_i,\: i = 1,2 $-dynamics, the second-oscillator type operators do not contribute. The energy levels are quantified by the formula:
\begin{equation}\label{eigval00}
E_{ n} = \hbar \omega^{*}\left(n + \frac{1}{2}\right).
\end{equation}

The wave function of the quantum Hilbert space are given by
$|\Psi \rangle = |n,m \rangle$.

\subsection{Representation in the quantum Hilbert space }

{ Without loss of generality, we restrict our developments to  the noncommutative quantum mechanics formalism  \cite{gouba-scholtz,ben-scholtz,a-hk} for the physical system of a harmonic oscillator.} We focus on the application of Hilbert-Schmidt
operators, bounded operators on the noncommutative classical configuration space,  denoted by
\beq
\mathcal H_c = \mbox{span}\left\{|n\rangle\  = \frac{1}{\sqrt{n !}}(a^{\dag})^{n}|0\rangle\right\}_{n=0}^{\infty}.
 \eeq
This space is isomorphic to the boson Fock space $\mathcal F = \{|n\rangle\}_{n=0}^{\infty}, $ where the annihilation and creation operators $a, a^{\dag}$ obey the Fock algebra $[a, a^{\dag}] = \mathbb I$.

The physical states of the system  represented on $\mathcal H_{q}$,  known  as the set of Hilbert-Schmidt operators,
is equivalent to the Hilbert space of square integrable functions, with
the classical configuration space $\mathcal H_c$, with  a general element of the quantum Hilbert space, in
''bra-ket'' notation given by
\beq
|\psi) = \sum_{n,m=0}^{\infty}c_{m,n}|m, n),
\eeq
with $ \left\{|m, n) := |m\rangle \langle n|\right\}_{m,n=0}^{\infty}$  a basis of $\mathcal H_{q}.$

Since the noncommutative configuration Hilbert space $\mathcal H_{c}$ is isomorphic to each of the  boson Fock spaces $\mathcal F_{\mathcal K}$ and $\mathcal F_{\mathcal P}$, respectively,  such that   $\mathcal F_{\mathcal K} \otimes \mathcal F_{\mathcal P}  =\mathfrak H = \mathcal H_{c} \otimes \mathcal H_{c}$, with $\mathfrak H$ given in (\ref{fock000}), the operators $\{\mathfrak{a}, \mathfrak{a}^{\dag}\}$ and $\{\mathfrak{b}, \mathfrak{b}^{\dag}\}$ are such that we obtain the following identifications:
\beq\label{hfock000}
\mathfrak{b} \otimes I_{\mathcal F_{\mathcal K}}\equiv \mathfrak{B}, \quad \mathfrak{b}^{\dag} \otimes I_{\mathcal F_{\mathcal K}}\equiv \mathfrak{B}^{\ddag},\quad I_{\mathcal F_{\mathcal P}} \otimes \mathfrak{a} \equiv \mathfrak{A},\quad   I_{\mathcal F_{\mathcal P}} \otimes \mathfrak{a}^{\dag} \equiv \mathfrak{A}^{\ddag},
\eeq
where the operators  $\{\mathfrak{A}, \mathfrak{A}^{\ddag}\}$ and $\{\mathfrak{B}, \mathfrak{B}^{\ddag}\}$ satisfy the commutators
\beq\label{opchi01}
[\mathfrak{A}, \mathfrak{A}^{\ddag}] = \1_q = [\mathfrak{B}, \mathfrak{B}^{\ddag}], \quad
[\mathfrak{A}, \mathfrak{B}^{\ddag}] = 0 = [\mathfrak{B}, \mathfrak{A}^{\ddag}], \quad [\mathfrak{A}, \mathfrak{B}] = 0,
\eeq
and have on $\mathcal H_{q}$ the  representations:
\beq\label{opchi03}
\mathfrak{B}|m, n)  &=&  \sqrt{m}|m-1, n)
\quad \quad  \mathfrak{B}^{\ddag}|m, n) = \sqrt{m + 1 }|m+1,  n),
\cr
\mathfrak{A}|m, n) &=&
\sqrt{ n}|m, n-1)
\quad \quad  \mathfrak{A}^{\ddag}|m, n) =  \sqrt{n + 1 } |m,  n+1).
\eeq
Then, as in (\ref{fock000}), we have
\beq\label{eig00}
|m, n)
= \frac{1}{\sqrt{m ! n !}}\left(\mathfrak{B}^{\ddag}\right)^{m}
\left(\mathfrak{A}^{\ddag}\right)^{n}|0\rangle \langle 0|
\eeq
where $\mathfrak{A}^{\ddag}$ may have an action on the right by $\mathfrak{A}$ on $|0\rangle \langle 0|$.
  $|||m, n)|| = 1$ and  $|0\rangle \langle 0|$
stands for the vacuum state on $\mathcal H_{q}$ (see for e.g. \cite{a-hk}).

\brmk
Within this framework, the states $|m, n)$, characterized by the quantum numbers $(m, n)$ for the two sectors, exhibit two distinct aspects. The "left" index $m$ corresponds to excitations in the $\mathcal{P}$-sector, which generates translations and is associated with the guiding center coordinates. Conversely, the "right" index $n$ corresponds to excitations in the $\mathcal{K}$-sector, which determines the energy through the Hamiltonian $\hat{H} \equiv \mathfrak{A}^{\ddag}\mathfrak{A}$.
\ermk

\section{Coherent states}
In this section, we construct explicit coherent states (CSs) for the exotic Landau problem and
rigorously verify their mathematical properties. Coherent states provide an essential bridge
between quantum and classical descriptions, offering insights into semiclassical behavior
while maintaining exact quantum characteristics.

With the help of the operators $\{\mathfrak{A},  \mathfrak{A}^{\ddag}\}$ satisfying (\ref{opchi01}),  the CSs related to
the Hamiltonian  $\hat{H}$
are  infinite component CSs \cite{ali-bagarello, aremua-gouba} denoted $|z, \bar{z}'; m)$  given
on $\mathcal{H}_q$ by
\begin{equation}
 |z, \bar{z}'; m) = |
 \bar{z}'; m\rangle  \langle \bar z|, \quad \mbox{where} \quad |\bar{z}'; m\rangle =
e^{-|z'|^2/2} \frac{\bar{z'}^{m}}{\sqrt{m!}}
|m\rangle
\end{equation}
 and
 \beq\label{CS000}
|z\rangle = e^{-|z|^2/2} \sum_{n = 0}^{\infty}\frac{z^{n}}{\sqrt{n!}}
|n\rangle
 \eeq
i.e.,
\begin{equation}\label{cohst00}
|z, \bar{z}'; m) = e^{-\left(|z|^2+ |z'|^2\right)/2}\bar{z'}^{m}
\sum_{n = 0}^{\infty} \frac{z^{{n}}}{\sqrt{m!n!}}|m, n), \;
m = 0, 1, 2, \dots, \infty.
\end{equation}
\subsection{Normalization to unity condition}
The normalization condition satisfied by the CSs $|z, \bar{z}'; m)$  given by
\begin{equation}\label{normal009}
\sum_{m = 0}^{\infty}( z, \bar{z}'; m|z, \bar{z}'; m)  = 1
\end{equation}
is obtained through the relations
\beq
\sum_{m = 0}^{\infty}( z, \bar{z}'; m|z, \bar{z}'; m)  = tr_{c}[(|z\rangle \langle z| )^{\dag}(|z\rangle \langle z| )] \sum_{m = 0}^{\infty} tr_{c}[(|\bar{z}'\rangle \langle \bar{z}'| )^{\dag}(|\bar{z}'\rangle \langle \bar{z}'| )]
\eeq
where
\beq
tr_{c}[(|z\rangle \langle z| )^{\dag}(|z\rangle \langle z| )]  = 1, \quad
\sum_{m = 0}^{\infty}tr_{c}[(|\bar{z}'\rangle \langle \bar{z}'| )^{\dag}(|\bar{z}'\rangle \langle \bar{z}'| )] = 1
\eeq
yielding (\ref{normal009}).

Next, let us verify that the constructed CSs (\ref{cohst00}) satisfy all Klauder's \cite{gazeauklauder} minimal requirements: (a) continuity in the labeling, (b) resolution
of unity, (c) temporal stability, and (d) action identity.
\subsection{ Continuity in  the labeling}\label{cont000}
\bpro\label{contin000}
This property consists in the following statement:
\beq
\forall z, z', z'' \in \C, \; |||z,\bar{z}';m) - |z',\bar{z}'',m)||^{2}_{\mathcal{HS}}\longrightarrow 0 \; \mbox{iff}\;
|z-z'|\longrightarrow 0\; \mbox{and}\; |\bar{z}'-\bar{z}''|\longrightarrow 0,
\eeq
where the norm $||.||_{\mathcal{HS}}$ is that of Hilbert-Schmidt.
\epro

 {\bf Proof.} See in the Appendix.

 $\hfill{\square}$
\subsection{Resolution of the identity}
\bpro\label{resolu000}
The CSs (\ref{cohst00}) satisfy  the following resolution of the identity
\begin{equation}\label{newresolv}
\frac{1}{\pi^{2}}\sum_{m = 0}^{\infty}\int_{\mathbb C^{2}}|z, \bar{z}'; m) (z, \bar{z}'; m|d^2zd^2z'
= \mathbb I_{q}
\end{equation}
where $ \mathbb I_{q}$ is the identity operator on $ \mathcal H_{q}$. The identity operator writes in terms of the states $|m, n)$   as follows:
\beq
\mathbb I_{q} &=& \sum_{m, n = 0}^{\infty}|m, n)(m, n| =  \sum_{m, n = 0}^{\infty}|m \rangle \langle n|
|n\rangle\langle m|.
\eeq
The identity operator on $ \mathcal H_{q}$ is  given by \cite{gouba-scholtz}
\beq{\label{resolv01}}
\mathbb I_{q} = \frac{1}{\pi}\int_{\mathbb C}dzd\bar{z}|z)e^{\overleftarrow{\partial_{\bar z}}\overrightarrow{\partial_{z}}} (z|.
\eeq
\epro
{\bf Proof.} Similar to the proof of Proposition 3.1 in \cite{a-hk}.

$\hfill{\square}$

\subsection{Temporal stability}
From the shifted  Hamiltonian $\mathbb{H} = \hat H  -  \frac{\hbar \omega^{*}}{2}\mathbb I_{\mathcal F_{\mathcal K}}$  with spectrum $\mathcal E_{n} =  \omega^{*} n,   \; \hbar = 1$, see (\ref{eigval00}), the dimensionless Hamiltonian denoted by ${\mathbb{H}}^{dim}$ is obtaned as ${\mathbb{H}}^{dim} = \frac{1}{\omega^{*}}{\mathbb{H}} $ with eigenvalues $e_{n} = n$, such that
\beq{\label{vect00}}
|z, \bar{z}'; m;\eta) &=& \mathbb U(\eta)|z, \bar{z}'; m), \quad
\mathbb U(\eta) = e^{-i  {\mathbb{H}}^{dim}    \eta}.
\eeq
Then, we have the following proposition:
 \bpro\label{timeev000}
Using  the  parameter $\eta$  introduced,  the states (\ref{cohst00})  fulfill the Klauder criterium
of temporal stability  relative to the classical time evolution operator
$\mathbb U(t)$:
\beq
\mathbb U(t)
|z, \bar{z}'; m;\eta)  =   e^{-i  {\mathbb{H}}^{dim}  t}|z, \bar{z}'; m;\eta)
=|z, \bar{z}'; m;\eta +t).
\eeq
\epro
{\bf Proof.}
Indeed, from (\ref{cohst00}) and (\ref{vect00}) together, we have
\beq
\mathbb U(t)
|z, \bar{z}'; m;\eta) &=& e^{-\left(|z|^2+ |z'|^2\right)/2}\bar{z'}^{m}
\sum_{n = 0}^{\infty} \frac{z^{{n}}}{\sqrt{m!n!}}e^{-i  e_{n}   \eta} e^{-i  {\mathbb{H}}^{dim}  t}|n, m) \cr
&=& e^{-\left(|z|^2+ |z'|^2\right)/2}\bar{z'}^{m}
\sum_{n = 0}^{\infty} \frac{z^{{n}}}{\sqrt{m!n!}}e^{-i  e_{n}   (\eta +t)} |n, m) \cr\cr
&=& |z, \bar{z}'; m;\eta +t).
\eeq

 $\hfill{\square}$
 
\subsection{Action identity}
 \bpro\label{actionid000}
The CSs $|z, \bar{z}'; m)$, given the shifted  Hamiltonian $\mathbb H = \hat H  -  \frac{\hbar \omega^{*}}{2}\mathbb I_{\mathcal F_{\mathcal K}}$  with spectrum $\mathcal E_{n} =  \omega^{*} n_,   \; \hbar = 1$, verify the action identity property
\beq
\sum_{m = 0}^{\infty}(z, \bar{z}'; m|\mathbb H|z, \bar{z}'; m) = \omega^{*} |z|.
\eeq
 \epro
 {\bf Proof.}
From the definition (\ref{cohst00}), we immediately get
\beq
\mathbb H|z, \bar{z}'; m)
= e^{-\left(|z|^2+ |z'|^2\right)/2}\bar{z'}^{m}
\sum_{n = 0}^{\infty} \frac{z^{{n}}\omega^{*} n|n,m)}{\sqrt{m!n!}}.
\eeq
Thereby
\beq
\sum_{m = 0}^{\infty}(z, \bar{z}'; m|\mathbb H|z, \bar{z}'; m)
&=& \omega^{*} e^{-\left(|z|^2+ |z'|^2\right)} \sum_{m = 0}^{\infty}\frac{|{z'}|^{2m}}{m!}
\sum_{n = 0}^{\infty} \frac{|z|^{2n} n}{n!}\cr
&=&\omega^{*} |z|.
\eeq

$\hfill{\square}$
\subsection{Density of probability}
This paragraph is devoted to  the sem-classical character of the CSs (\ref{cohst00}) by examining how they do evolve in time under the action of the time evolution
operator from the physical Hamiltonian describing the quantum system.

From the definition of the CSs $|z, \bar{z}'; m)$, we have the following overlap
\beq\label{overlap007}
(z, \bar{z}'; m|z_0, \bar{z}'; m) &=& e^{-|z'|^2}\frac{|z'|^{2m}}{m!}e^{-\left(|z|^2+ |z_0|^2\right)/2}e^{z_0 \bar z},
\eeq
such that given a  normalized state $|z_0, \bar{z}'; m)$, we define the density of probability as
\beq
z \mapsto \varrho_{{z_0}}(z) \deq   |(z, \bar{z}'; m|z_0, \bar{z}'; m)|^2
=  \left\{e^{-|z'|^2}\frac{|z'|^{2m}}{m!}\right\}^{2}\frac{e^{z_0 \bar z +  z\bar z_0}}{e^{\left(|z|^2+ |z_0|^2\right)}}.
\eeq
The time evolution behavior of $\varrho_{z_0}(z)$ is provided from  the shifted  Hamiltonian $\mathbb H = \hat H  -  \frac{\hbar \omega^{*}}{2}\mathbb I_{\mathcal F_{\mathcal K}}$   with spectrum $\mathcal E_{n} =  \omega^{*} n,   \; \hbar = 1$ by
\beq
z \mapsto \varrho_{z_0}(z,t) \deq |(z, \bar{z}'; m|e^{-i  {\mathbb{H}} t}|z_0, \bar{z}'; m)|^2,
\eeq
where $z_0(t)\deq z_0e^{-i\omega^{*}t}$, suggesting  pure rotation in phase space,  with $|z_0(t)| = |z_0| $. Thereby,
\beq
\varrho_{z_0}(z,t) \deq |(z, \bar{z}'; m|e^{-i  {\mathbb{H}} t}|z_0, \bar{z}'; m)|^2 =  \left\{e^{-|z'|^2}\frac{|z'|^{2m}}{m!}\right\}^{2}\frac{e^{z_0(t) \bar z +  z\bar z_0(t)}}{e^{\left(|z|^2+ |z_0(t)|^2\right)}},
\eeq
or in terms of Meijer-G functions:
\begin{eqnarray}\label{tempdensmeijer000}
    \varrho_{z_0}(z,t)
		&=&
		\left\{e^{-|z'|^2}\frac{|z'|^{2m}}{m!}\right\}^2
		\frac{G^{1,0}_{0,1}\left(
		-\bar z_0(t)z
		\Big|
		\begin{matrix}
			0
		\end{matrix}
		\right)
		G^{1,0}_{0,1}\left(
		-z_0(t)\bar z
		\Big|
		\begin{matrix}
			0
		\end{matrix}
		\right)}{G^{1,0}_{0,1}\left(
		-(|z|^2+|z_0(t)|^2)
		\Big|
		\begin{matrix}
		0
		\end{matrix}
		\right)}.
		\end{eqnarray}
%
%
%
\begin{center}
	\begin{figure}[h]
			\begin{subfigure}[b]{0.31\textwidth}
				\includegraphics[width=\textwidth]{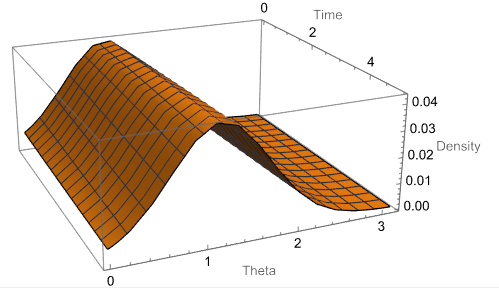}
				\caption{$m=2$}
		\end{subfigure}
		\begin{subfigure}[b]{0.31\textwidth}
			\includegraphics[width=\textwidth]{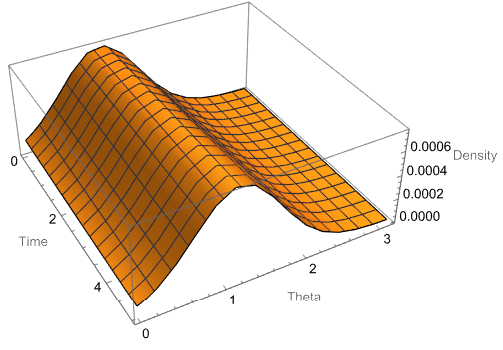}
			\caption{$m=5$}
		\end{subfigure}
		\begin{subfigure}[b]{0.31\textwidth}
			\includegraphics[width=\textwidth]{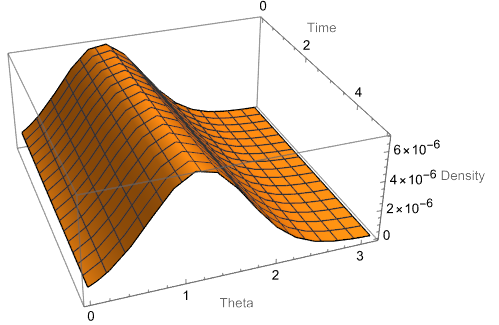}
		\caption{$m=7$}
	\end{subfigure}
		\caption{\it \small Plots of the temporal density of probability $\varrho_{z_0}(z,t)$  \ref{tempdensmeijer000}: (a): for $m=2$; (b): for $m=5$; (c): for $m=7$, as a function of the angle $\theta \in [0, \pi]$, argument of the complex nmber $z  = |z| e^{i\theta}$,  and the time $t \in [0, 5]$ (in normalized units).
			}
	\end{figure}
\end{center}
Figure 1 presents the probability density \(\rho_{z_0}(z,t)\), which encodes the quantum interference between two CSs evolving under the quantum Hamiltonian. The analysis reveals interesting physical insights as demonstrated across the three graphs, which show a systematic decrease in density magnitude across several orders of magnitude: from \(\rho_{\max} \approx 0.04\) in Fig. 1(a), to \(\rho_{\max} \approx 6 \times 10^{-4}\) in Fig. 1(b), and finally to \(\rho_{\max} \approx 6 \times 10^{-6}\) in Fig. 1(c). This decay illustrates a scaling behavior characteristic of quantum mechanical decay or dissipative processes. As a phase space distribution, it is important to note that since \(z\) and \(\bar{z}\) represent complex coordinates in quantum phase space, by fixing \(z = \frac{1}{\sqrt{2}}(x + ip)\) where \(x\) is position and \(p\) is momentum  the probability density \(\rho_{z_0}(z,t)\) essentially acts as a Husimi Q-function or a CS representation of the quantum state. This representation is closely related to the Wigner function \cite{glauber-schleich}.
For \(m = 2\), the probability density exhibits relatively simple oscillatory behavior, suggesting that the system behaves in a way that is closest to classical intuition. The observed oscillations correspond to the classical periodic motion of a harmonic oscillator. However, as \(m\) increases, the phase space distribution becomes more delocalized.  Consequently, the parameter \(m\) effectively controls the degree of quantum complexity of the CSs dynamics.
\subsection{Nonclassical behavior}
Let us verify in this paragraph that the CSs $|z, \bar{z}'; m)$ display statistical aspect.
We have from the definition (\ref{cohst00}), the following algebra
\beq\label{pnd000}
|(m, n|z, \bar{z}'; m)|^2 &=& (m, n|z, \bar{z}'; m)(     z, \bar{z}'; m|m, n) \cr
&=& \left\{e^{-\left(|z|^2+ |z'|^2\right)/2}\bar{z'}^{m}
\sum_{k = 0}^{\infty} \frac{z^{{k}}}{\sqrt{k!m!}}\delta_{nk}\right\} \left\{e^{-\left(|z|^2+ |z'|^2\right)/2}{z'}^{m}
\sum_{p = 0}^{\infty} \frac{\bar z^{{p}}}{\sqrt{m!p^!}}\delta_{np}\right\}\cr
&=& e^{- |z'|^{2} }\frac{|z'|^{2m}}{m!} e^{- |z|^{2} }\frac{|z|^{2n}}{n !}
\eeq
which displays that the CSs $|z, \bar{z}'; m)$ obey the photon-number Poisson distribution corresponding to
a Mandel parameter $\mathcal Q = 0$ \cite{mandel-wolf}, which can be interpreted as an absence of  quantum correlations between photons (classical behavior)
such that the states minimize the Heisenberg uncertainty with fluctuations compatible with quantum mechanics. In the context of the exotic Landau problem,  these photons can be associated with excitation quanta of the harmonic oscillators associated with the conserved quantities $\hat{\mathcal{K}}_i, \hat{\mathcal{P}}_i$.
  \begin{center}
	\begin{figure}[h]
	\begin{subfigure}[b]{0.31\textwidth}
		\includegraphics[width=\textwidth]{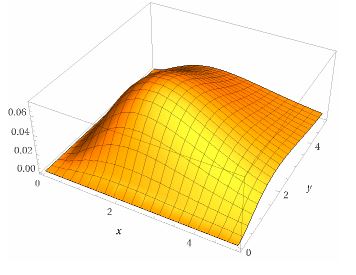}
		\caption{$m=2$}
	\end{subfigure}
	\begin{subfigure}[b]{0.31\textwidth}
		\includegraphics[width=\textwidth]{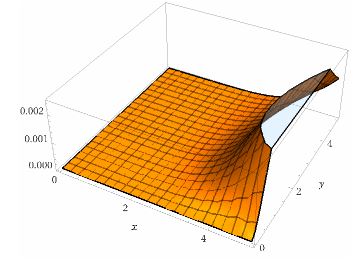}
		\caption{$m=5$}
	\end{subfigure}
		\begin{subfigure}[b]{0.31\textwidth}
			\includegraphics[width=\textwidth]{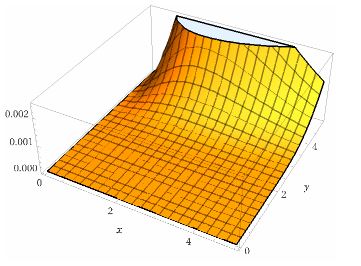}
			\caption{$m=7$}
		\end{subfigure}
		\caption{\it \small Plots of the Photon Number Distribution (PND) (\ref{pnd000})  versus $x = |z|$ and $y= |z'|$: (a) for $m = 2$ and $n = 2$; (b) for $m = 2$ and $n = 10$; (c) for $m = 10$ and $n = 2$.
		}
	\end{figure}
\end{center}
Figure 2 illustrates the photon number distribution (PND) associated with the CSs \( |z, \bar{z}'; m) \), as detailed in Eq. (49). The distribution demonstrates a factorization into Poisson laws for the different sectors, which emphasizes the classical-like character of the constructed CSs. Panels (a)-(c) depict the distributions for varying values of the chiral quantum number \( m \) and excitation number \( n \). As these parameters change, the distributions maintain a Poissonian profile, characterized by differing widths and amplitudes. Notably, the statistics remain strictly Poissonian across all scenarios, indicated by a vanishing Mandel parameter.

\brmk\label{rem000}
The thermodynamics cannot be investigated here when taking both $n, m$ sector contributions into account. Indeed,
	the eigenvalues (\ref{eigval00}) are such that the Hamiltonian is infinitely degenerate in the $m$ chiral sector. Then, starting from
 \begin{equation}
  (z, \bar{z}'; m|\rho|z, \bar{z}'; m)
= (z, \bar{z}'; m|\left\{\frac{1}{Z}\sum_{m, n=0}^{\infty}e^{-\beta {\hat H}}|m, n)  (m, n|
\right\}|z, \bar{z}'; m)
 \end{equation}
 provides
 \begin{equation}
Z =\mbox{Tr}e^{-\beta {\hat H}_{-}}=
\sum_{n=0}^{\infty}\langle n|e^{-\beta\hbar \omega^{*}\left(n + \frac{1}{2}\right) }|n\rangle
\sum_{m=0}^{\infty}\langle m|m\rangle
 \end{equation}
leading, because of the infinite sum $\displaystyle\sum_{m =0}^{\infty}\langle m|m\rangle$,
 to  a misconstruction of the partition function $Z$.
\ermk
\subsection{Free particle propagator}
This paragraph deals with the free particle propagator derivation from the resolution of the identity property provided by the constructed CSs (\ref{cohst00}), by following the methods developed in \cite{hstan, gangopadhyay-scholtz}. As expected properties, the CSs  displayed an ultra-violet cutoff, with the overlap between two CSs bringing a  transition amplitude of Gaussian type.

Before proceeding further,  introduce  a set of
dimensionless complex variables
\beq\label{moment001}
p = \sqrt{\frac{1-eB\theta}{2eB}}[p^{1} + i p^{2}], \qquad
\bar p  = \sqrt{\frac{1-eB\theta}{2eB}}
[p^{1} - i p^{2}],
\eeq
where the $ p^{i},  \; i=1,2,$  satisfy in the momentum representation the following equality $\hat P^{i}|p) = p^{i} |p)$, with the orthonormalisation and completeness relations delivered as follows
\beq\label{presolv}
(p'|p) = \delta(p - p'),\qquad \int d^{2} p |p) (p| = \mathbb{I}_{q}.
\eeq
Next, consider  in the momentum representation  the vector $|p)$ given by \cite{gangopadhyay-scholtz}
\beq
|p) = \sqrt{\frac{\theta}{2\pi \hbar^{2}}}e^{\frac{i}{\hbar}\sqrt{\frac{\theta}{2}}(\bar p \mathfrak{A}
 + p \mathfrak{A}^{\ddag})}
\eeq
and the wave function of the CSs $|z, \bar z)$ in  this basis  given by
\beq\label{presolv00}
(z, \bar z|p)  = \sqrt{\frac{\theta}{2\pi\hbar^{2}}}
e^{-\frac{\theta}{4\hbar^{2}} |p|^{2}}
e^{\frac{i}{\hbar}\sqrt{\frac{\theta}{2}}(p\bar z + \bar p z)}.
\eeq
Let us fix the Hamiltonian
$\hat H(\hat P)  =
\frac{\hat P^{2}}{2M}$ (see Eq.(\ref{jjjj})) with eigenvalues  $\frac{p^{2}}{2M}$
on the eigenstates $|p)$
representation,  where from (\ref{moment001}) we get
$|p|^{2} =
\frac{1-eB\theta}{2eB}[(p^{1})^{2} + (p^{2})^{2}] \equiv
\frac{1-eB\theta}{2eB} |\tilde p|^{2}$.

Then, the propagator over a small segment expresses as follows,
\beq
(z^{i+1}; m|e^{-i \tau \hat H}|z^{i}; m)
   &=&  \int d^{2} p_i \,(z^{i+1}; m|e^{-i \tau  \hat H} |p_i) (p_i|z^{i}; m)
   \cr
&=&
\left[\frac{1-eB\theta}{2eB}\right] \left[\frac{\theta}{2\pi \hbar^{2}}\right]
\int d^{2} \tilde p^i \,e^{-i \tau \frac{(1-eB\theta)|\tilde p^i|^2}{4MeB}} \cr
&& \left\{e^{-\frac{\theta}{4\hbar^{2}}\frac{1-eB\theta}{2eB}
 |\tilde p^i|^{2}} e^{\frac{i}{\hbar}\sqrt{\frac{\theta}{2}}\sqrt{\frac{1-eB\theta}{2eB}}(\tilde p^i\bar  z^{i+1} +
 \bar{\tilde p}^i z^{i+1})}\right\}\cr
 && \times
  \left\{e^{-\frac{\theta}{4\hbar^{2}}\frac{1-eB\theta}{2eB} |\tilde p^i|^{2}}
e^{-\frac{i}{\hbar}\sqrt{\frac{\theta}{2}}\sqrt{\frac{1-eB\theta}{2eB}}(\tilde p^i\bar z^{i} + \bar{\tilde p}^i z^{i})} \right\}
\cr
&=& \left[\frac{M\theta}{\frac{2eMB\theta}{1-eB\theta} + i\tau } \right]
e^{-\frac{2M\theta}{2M\theta + i\tau \left(\frac{1}{eB} - \theta \right)}|z^{i+1}-z^{i}|^{2}}.
\eeq
The relations (\ref{resolv01}) and (\ref{presolv}) allow us to write down the path integral for the
free particle propagation kernel \cite{gangopadhyay-scholtz} on the two-dimensional
noncommutative space. We have the following:
\bpro\label{propag000}
From
the resolution of the identity (\ref{resolv01}), we get
\beq
(z^{f}, t_f|z^{0}, t_0)& := & \lim_{n\rightarrow \infty}\int
\frac{1}{\theta} \left(\frac{1}{\pi^{2}}\right)^{n}
\left(\prod^{n}_{j=1} d^{2} z^{j} \right)(z^{f}, t_f|z^{n}, t_n)
\star_{n} (z^{n}, t_n|\dots |z^{1}, t_1)\cr &&\star_{1}
(z^{1}, t_1|z^{0}, t_0) \nonumber
\\
\eeq
where the  product $\star_{j} $ is given as
\beq
\star_{j}  = e^{\overleftarrow{\partial_{\bar{z_j}}}\overrightarrow{\partial_{z_j}}}  = \int \frac{d^2 v}{\pi}
e^{- |v|^{2}}
e^{\bar v\overleftarrow{\partial_{\bar z_j}} + v\overrightarrow{\partial_{z_j}}}.
\eeq
Thus, we arrive at the following relation
\beq
\int d^{2}z^{i}   \, (z^{i+1}, t_{i+1}|z^{i}, t_{i})
 \star_{i}
(z^{i}, t_{i}|z^{0},  t_{0}) =   \frac{ \pi M\theta}{\frac{2eMB\theta}{1-eB\theta} +  2i\tau }
e^{-\frac{2 M\theta}{2M\theta + 2i\tau \left(\frac{1}{eB} - \theta \right)}|z^{i+1} - z^{0}|^{2}}.
\eeq

\epro
{\bf Proof.}  Indeed, we have
\beq
&&\int d^{2}z^{i}   \, (z^{i+1}, t_{i+1}|z^{i}, t_{i})
 \star_{i}
(z^{i}, t_{i}|z^{0},  t_{0})\cr
&=& N_1 N_2 \int d^{2}z^{i} \, \left\{ \int  \frac{d^{2} v }{\pi} \, e^{-|v|^{2}}
e^{-\beta_1|z^{i+1} - z^{i}|^{2}}  e^{\bar v\overleftarrow{\partial_{ \bar z^{i}_{-}}}+v\overrightarrow{\partial_{z^{i}}}}
 e^{-\beta_2|z^{i} - z^{0}|^{2}}\right\} \cr
&=& \frac{N_1 N_2}{\beta_1}\int d^{2} v \, e^{-|v|^{2}} \left\{ e^{-\gamma|v|^{2}}
e^{\beta_2|v|^{2}} e^{-\sqrt{\beta_1 \gamma}
[v (\bar z^{i+1} - \bar z^{0}) + \bar v (z^{i+1} - z^{0}))]}\right\}\cr
&=& \frac{N_1 N_2}{\beta_1}\frac{\pi}{\Lambda}e^{-\frac{\beta_1 \gamma}{\Lambda}|z^{i+1} - z^{0}|^{2}}, \quad \Lambda = 1+\gamma-  \beta_2, \quad \gamma = \beta_2/\beta_1,
 \eeq
with Gaussian integral of the type $\displaystyle \int d^2\bar{p} e^{-A|\bar{p}|^2 + B\bar{p} + \bar{B}\bar{\bar{p}}} = \frac{\pi}{A} e^{\frac{|B|^2}{A}}$ used,  $N_1 =  N_2 =  \sqrt{\frac{1-eB\theta}{2eB}}\Theta$ and
$\beta_1 = \beta_2 = \Theta$, where  $\displaystyle \Theta = \frac{2M\theta}{2M\theta + i \tau \left(\frac{1}{eB} - \theta \right)}, \gamma =1, \Lambda = 2-\beta$,
 such that  $\displaystyle \frac{N_1 N_2 \pi}{\beta_1  \Lambda} = \frac{ \pi M\theta}{\frac{2eMB\theta}{1-eB\theta} +  2i\tau }$ and $\frac{\beta_1 \gamma}{\Lambda}  = \frac{2 M\theta}{2M\theta + 2i\tau \left(\frac{1}{eB} - \theta \right)}$.
Thereby
\beq
\int d^{2}z^{i}   \, (z^{i+1}, t_{i+1}|z^{i}, t_{i})
 \star_{i}
(z^{i}, t_{i}|z^{0},  t_{0}) =   \frac{ \pi M\theta}{\frac{2eMB\theta}{1-eB\theta} +  2i\tau }
e^{-\frac{2 M\theta}{2M\theta + 2i\tau \left(\frac{1}{eB} - \theta \right)}|z^{i+1} - z^{0}|^{2}}.
\eeq

$\hfill{\square}$

Assuming that
\beq
&& \lim_{n\rightarrow \infty}\int \frac{1}{\theta}\left(\frac{1}{\pi^{2}}\right)^{n-1}
\left(\prod^{n-1}_{j=1}d^{2} z^{j}\right)  (z^{n}, t_n|z^{n-1}, t_{n-1})
\star_{n-1} (z^{n-1}, t_{n-1}|\dots |z^{1}, t_1) \cr
&&\star_{1}
(z^{1}, t_1|z^{0}, t_0)\cr
&=& \lim_{n\rightarrow \infty}\frac{M}{\frac{2eMB\theta}{1-eB\theta} + in\tau }
  e^{-\frac{2M\theta}{2M\theta + in\tau \left(\frac{1}{eB} - \theta \right)}|z^{n}- z^{0}|^{2}},
\eeq
we obtain, after some algebra,
\beq\label{propag007}
&& \lim_{n\rightarrow \infty}\int \frac{1}{\theta}\left(\frac{1}{\pi^{2}}\right)^{n}
\left(\prod^{n}_{j=1} d^{2} z^{j} \right) (z^{f}, t_f|z^{n}, t_{n})
\star_{n} (z^{n}, t_{n}|\dots |z^{1}, t_1) \star_{1}
(z^{1}, t_1|z^{0}, t_0)\cr
  &=& \left[\frac{1-eB\theta}{2eB\theta}\right] \left[\frac{2M\theta}{2M\theta + iT \left(\frac{1}{eB} - \theta \right)}\right]
  e^{-\frac{2M\theta}{2M\theta + iT \left(\frac{1}{eB} - \theta \right)}|z^{f}- z^{0}|^{2}},  (n + 1)\tau = T = t_f - t_0.
\eeq
Note that    (\ref{propag007}) is analogue to the one obtained in \cite{gangopadhyay-scholtz} given by
\beq
(z_{f}, t_f|z_{0}, t_0) = \frac{m}{m\theta + iT} \exp \left\{-\frac{m}{m\theta + iT} (\overrightarrow x_f - \overrightarrow x_0)^{2}\right\}
\eeq
highlighting   the ultra-violet cutoff induced by
the noncommutative parameter. Moreover, taking the limit $T = t_f-t_0\rightarrow 0$ of (\ref{propag007}), we recover  the following expression:
\beq
\lim_{T\rightarrow 0}(z^{f}, t_f|z^{0}, t_0) &=& \left[\frac{1-eB\theta}{2\theta eB}\right]
 e^{-|z^{f}- z^{0}|^{2} } \cr
&\propto& (z^{f}|z^{0})
\eeq
with
$ (z^{f}|z^{0})=
 e^{-|z^{f}- z^{0}|^{2} }$
being the expected  Gaussian  transition amplitude between two CSs, see paragraph \ref{cont000}.

\section{Vector coherent states  construction in a noncommutative Hilbert space}
In this section, we discuss two classes of vector coherent states (VCSs) from the constructed   CSs \ref{cohst00} following the scheme developed in \cite{ali-bagarello, a-hk}.
We also investigate their  main mathematical properties and  their
physical insights.
\subsection{The setup}
Let $\mathcal M_{2}(\mathbb C)$, the space of $2 \times 2$  complex  matrices,  be  a
locally compact space equipped with a measure $d\mu$   as the parameter space  defining the VCSs.
Consider the quantum Hilbert space  $\mathcal H_{q}$  of Hilbert-Schmidt operators
acting on the noncommutative configuration (Hilbert) space  $\mathcal H_{c} = span\{|n\rangle, n \in \mathbb N\}$.
Let $F_{n}(\mathfrak Z): \mathcal M_{2}(\mathbb C)\rightarrow \mathcal B(\mathcal H_{c})$,  where $\mathcal B(\mathcal H_{c})$ is
the set of bounded
operators
on $\mathcal H_{c}$,  be a set of continuous mappings satisfying the conditions:
\bitem
\item [(i)]
for each $\mathfrak Z
\in \mathcal M_{2}(\mathbb C)$, the following normalization condition
\beq{\label{bound00}}
0 < \mathcal N(\mathfrak Z) = \sum_{n \in \mathbb N} tr_{c} [|F_{n}(\mathfrak Z)|^{2}] < \infty
\eeq
 is satisfied, where $tr_{c}$ stands for the trace over $\mathcal H_{c}$ and
$|F_{n}(\mathfrak Z)|^{2} = \left[F_{n}(\mathfrak Z)F_{n}(\mathfrak Z)^{*}\right]^{1/2}$ denotes the positive part of the operator
$F_{n}(\mathfrak Z)$;
\item [(ii)] for each $\mathfrak Z
\in \mathcal M_{2}(\mathbb C)$, there exists a bounded linear map $T(\mathfrak Z): \mathbb C^{2} \rightarrow \mathbb C^{2} \otimes \mathcal H_{c}$
defined by
\beq
T(\mathfrak Z)\chi^{j}  = (\mathcal N(\mathfrak Z))^{-1/2} \sum_{n \in \mathbb N} F_{n}(\mathfrak Z)
|\chi^{j}, n \rangle \qquad j=1,2;
\eeq
\item [(iii)] for each $\mathfrak Z
\in \mathcal M_{2}(\mathbb C)$, the following expression given on $\mathbb C^{2} \otimes \mathcal H_{c}$ holds:
\beq{\label{bound009}}
F_{n}(\mathfrak Z)|\chi^{j}, \tilde n\rangle &=&
\frac{\mathfrak Z^{n} \bar{\mathfrak Z}^{\tilde n} }{\sqrt{R(n)R(\tilde n)}}|\chi^{j}, \tilde n\rangle
\eeq
where $
\mathfrak Z = diag(z_{1}, z_{2}),\,   z_{j} = r_{j}e^{i \theta_{j}}$
with $r_{j} \geq 0, \theta_{j} \in [0,2\pi)$ and  $R(n) = n !\mathbb I_{2}$.
 \eitem
\subsection{Vector coherent states construction}
With the above  setup, from (\ref{bound009}),  and following the construction
 provided above, based in \cite{a-hk},
the set of vectors  formally given by
\beq{\label{ncvcs00}}
|\mathfrak Z, \bar{\mathfrak Z'}, \eta,  j,  n, m, \tilde n, \tilde m)
&=& (\mathcal N(\mathfrak Z, \bar{\mathfrak Z'}))^{-1/2}\sum_{m, n=0}^{\infty}
\frac{\mathfrak Z^{n} \bar{\mathfrak Z'}^{m} }{\sqrt{R(n)R(m)}}
\frac{\bar{\mathfrak Z}^{\tilde n}\mathfrak Z'^{\tilde m}}{\sqrt{R(\tilde n)R(\tilde m)}} e^{-i \eta \mathcal E_{n}}\cr
 &&\times |\chi^{j}\rangle \otimes |\tilde n \rangle   \langle \tilde m| \otimes |m\rangle \langle n|
\eeq
where $
\mathfrak Z = diag(z_{1}, z_{2}),\,   z_{j} = r_{j}e^{-i \theta_{j}}$
with $r_{j} \geq 0, \theta_{j} \in [0,2\pi)$, and $\bar{\mathfrak Z'} = diag(\bar{z}'_{1}, \bar{z}_{2}',), \bar{z}'_{j} = \rho_{j}e^{i \varphi_{j}}$ with $\rho_{j} \geq 0, \varphi_{j} \in [0,2\pi)$,  and
$R(m) = m!\mathbb I_{2},  R(\tilde m) = \tilde m!\mathbb I_{2},   R(n) = n!\mathbb I_{2},  R(\tilde n) = \tilde n!\mathbb I_{2}$,
forms a set of VCSs on  $\mathbb C^{2} \otimes \mathcal H_{q} \otimes \mathcal H_{q}$.

The normalization condition to unity   given by
\beq\label{normnvcs00}
\sum_{j=1}^{2}\sum_{\tilde m, \tilde n = 0}^{\infty}
(\mathfrak Z, \bar{\mathfrak Z'}, \eta,  j,  n, m, \tilde n, \tilde m|\mathfrak Z, \bar{\mathfrak Z'}, \eta,  j,  n, m, \tilde n, \tilde m) = 1
\eeq
of the VCSs (\ref{ncvcs00}) yields
\beq
\mathcal N(\mathfrak Z, \bar{\mathfrak Z'}) =
e^{2(r^{2}_{1} + \rho^{2}_{1})}  + e^{2(r^{2}_{2} + \rho^{2}_{2})}.
\eeq
Let
$D = \{(z_{1}, z_{2})  \in \mathbb C^2  \,| \; \,  |z_{j}|< \infty, j=1,2 \}$,
$\mathcal D =
\{(z'_{1}, z'_{2})  \in \mathbb C^2  \,| \; \,  |z'_{j}|< \infty, j=1,2 \}$. Then, we have
\bpro\label{resolu001}
The VCSs (\ref{ncvcs00})  satisfy on the quantum Hilbert space $\mathbb C^{2} \otimes \mathcal H_{q} \otimes \mathcal H_{q}$ a resolution of the
identity as follows:
\beq{\label{ncres03}}
&&\sum_{j=1}^{2}\sum_{\tilde m = 0}^{\infty}\sum_{\tilde n = 0}^{\infty}\frac{1}{\tilde m! \tilde n!}  \int_{D \times \mathcal  D}
d\mu(\mathfrak Z, \bar{\mathfrak Z'})
(\overrightarrow{\partial_{z_{j}}})^{\tilde m }
(\overrightarrow{\partial_{\bar z'_{j}}})^{m}
[\mathcal N(\mathfrak Z, \bar{\mathfrak Z'})\cr
&&|\mathfrak Z, \bar{\mathfrak Z'}, \eta,  j,  n, m, \tilde n, \tilde m)
(\mathfrak Z, \bar{\mathfrak Z'}, \eta,  j,  n, m, \tilde n, \tilde m|]
(\overleftarrow{\partial_{\bar z_{j}}})^{\tilde m}(\overleftarrow{\partial_{z'_{j}}})^{m}
= \mathbb I_{2} \otimes \mathbb I_{q} \otimes \mathbb I_{q}
\eeq
where the measure $d\mu(\mathfrak Z, \bar{\mathfrak Z'})$ is given on  $D \times \mathcal D$  by
\beq{\label{ncres05}}
d\mu(\mathfrak Z, \bar{\mathfrak Z'}) =   \frac{1}{(2\pi)^{2}}
\prod_{j=1}^{2}\lambda(r_{j}) \varpi(\rho_{j})dr_{j}d\rho_{j}
d\theta_{j}d\varphi_{j}.
\eeq
\epro
{\bf Proof.} Similar to the proof of Proposition 4.2 in \cite{a-hk}.

$\hfill{\square}$

\brmk
From the definition of the VCSs $|\mathfrak Z, \bar{\mathfrak Z'}, \eta,  j,  n, m, \tilde n, \tilde m)$, proceeding as in the case of  Propositions (\ref{timeev000}) and (\ref{actionid000}), it is straightforward to see that the temporal stability and  action identity properties given by
\beq
\mathbb U(t)|\mathfrak Z, \bar{\mathfrak Z'}, \eta,  j,  n, m, \tilde n, \tilde m)  = |\mathfrak Z, \bar{\mathfrak Z'}, \eta + t,  j,  n, m, \tilde n, \tilde m),
\quad \mathbb U(t) := e^{-i t \mathbb H},
\eeq
\beq
\sum_{j=1}^{2}\sum_{\tilde n, \tilde m = 0}^{\infty}
(\mathfrak Z, \bar{\mathfrak Z'}, \eta,  j,  n, m, \tilde n, \tilde m|\mathbb H|\mathfrak Z, \bar{\mathfrak Z'}, \eta,  j,  n, m, \tilde n, \tilde m)  =  \omega_c|\mathfrak Z|^{2}.
\eeq
are satisfied.
\ermk

\subsection{Quaternionic vector coherent states}
This part is devoted to the quaternionic extension of the CSs (\ref{cohst00}), known  as quaternionic vector coherent states (QVCSs),  on the  Hilbert space  $\mathbb C^{2} \otimes \mathcal  H_{q} \otimes \mathcal  H_{q}$. First, we achieve the completeness relation verified by these  QVCSs and then
derive, and analyze,  the  uncertainty relations specific to their representation  and their dynamical evolution.
\subsubsection{Construction}
We briefly discuss now the QVCSs construction and their  connection with  the studied  VCSs.
In (\ref{ncvcs00}),
set $\mathfrak Z = diag(z,\bar z)$ and $\bar{\mathfrak Z'}  = diag(z', \bar z')$ where
$z = re^{-i \tilde \phi},  \bar z' = \rho e^{i \tilde \varphi}$ with $r, \rho \geq 0, \, \tilde \phi, \tilde \varphi \in [0, 2\pi)$.
Consider  $u, v \in SU(2)$ and take $\mathcal Z = U \mathfrak Z U^{\dag}, \bar{\mathcal Z'} = V \bar{\mathfrak Z'}  V^{\dag}$ where
$U = diag(u,u), \, V = diag(v,v)$.

Next, introduce the quaternions
$\mathfrak Q = A(r)e^{i \vartheta \Theta(\hat n )}$, and $ \mathfrak Q' = B(\rho)e^{i \gamma \tilde \varTheta(\hat k) }$ with
$\varTheta(\hat n ) = diag(\sigma(\hat n ), \sigma(\hat n )) , \,
\tilde\varTheta(\hat k ) = diag(\tilde \sigma(\hat k ), \tilde \sigma(\hat k ))$,  where $A(r) = r\mathbb I_{2},
\, B(\rho) = \rho\mathbb I_{2}$ and
\beq\label{quat007}
\sigma(\hat n ) = \left(\begin{array}{cc}
\cos{\phi} & e^{i \eta}\sin{\phi} \\
e^{-i \eta}\sin{\phi} & -\cos{\phi}
                        \end{array}
\right) \qquad \tilde \sigma(\hat k) = \left(\begin{array}{cc}
\cos{\varphi} & e^{i \varrho}\sin{\varphi} \\
e^{-i \varrho}\sin{\varphi}  & -\cos{\varphi}
                                             \end{array}
\right)
\eeq
where $\phi, \varphi \in [0, \pi]$ and $\vartheta, \gamma, \eta,\varrho \in [0, 2\pi)$.

From the scheme developed in  \cite{mural-santhar, a-hk}, since $u, v$ are given as $u = u_{\xi_{1}}u_{\phi_{1}}u_{\xi_{2}}, \,
v = v_{\zeta_{1}}v_{\phi_{2}}v_{\zeta_{2}}$ with $ u_{\xi_{1}} = diag(e^{i \xi_{1}/2}, e^{-i \xi_{1}/2}), \,
u_{\xi_{2}} = diag(e^{i \xi_{2}/2}, e^{-i \xi_{2}/2}), \,  v_{\zeta_{1}} = diag(e^{i \zeta_{1}/2}, e^{-i \zeta_{1}/2}),
 v_{\zeta_{2}} = diag(e^{i \zeta_{2}/2}, e^{-i \zeta_{2}/2})$, and
\beq
u_{\phi_{1}} = \left(\begin{array}{cc}
\cos{\frac{\phi_{1}}{2}} & i \sin{\frac{\phi_{1}}{2}} \\
i \sin{\frac{\phi_{1}}{2}} & \cos{\frac{\phi_{1}}{2}}
                     \end{array}
\right) \quad v_{\phi_{2}} = \left(\begin{array}{cc}
\cos{\frac{\phi_{2}}{2}} & i \sin{\frac{\phi_{2}}{2}} \\
i \sin{\frac{\phi_{2}}{2}} & \cos{\frac{\phi_{2}}{2}}
                     \end{array}
\right), \;\; \xi_{1}, \xi_{2}, \zeta_{1}, \zeta_{2} \in [0,2\pi)
\eeq

for $ \xi_{1}= \xi_{2} = \eta$ and $\zeta_{1}=  \zeta_{2} = \varrho$, we get the following identifications:
$\mathcal Z = r(\mathbb I_{2}\cos{\vartheta} + i
\varTheta(\hat n )\sin{\vartheta}) = \mathfrak Q, \,
\mathcal W = \rho(\mathbb I_{2}\cos{\gamma} + i \tilde\varTheta(\hat k )\sin{\gamma}) = \mathfrak Q'$.

{Thereby, the QVCSs are given by  $|U\mathfrak Z U^{\dag}, V \mathfrak  W V^{\dag}, \tau, j,  n, m, \tilde n, \tilde m) =
|\mathfrak Q, \mathfrak Q',  \tau, j,  n, m, \tilde n, \tilde m)$ such that }
\beq{\label{nqvcs00}}
|\mathfrak Q, \bar{\mathfrak Q'}, \eta,  j,  n, m, \tilde n, \tilde m)
&=& (\mathcal N(r, \rho))^{-1/2}\sum_{m, n=0}^{\infty}
\frac{\mathfrak Q^{n} \bar{\mathfrak Q'}^{m} }{\sqrt{R(n)R(m)}}
\frac{\bar{\mathfrak Q}^{\tilde n}\mathfrak Q'^{\tilde m}}{\sqrt{R(\tilde n)R(\tilde m)}} e^{-i \eta \mathcal E_{n}}\cr
 &&\times |\chi^{j}\rangle \otimes |\tilde n \rangle   \langle \tilde m| \otimes |m\rangle \langle n|
\eeq
They satisfy a normalization  condition to unity given by
\beq
\sum_{j=1}^{2}\sum_{\tilde n, \tilde m = 0}^{\infty}
(\mathfrak Q, \bar{\mathfrak Q'}, \eta,  j,  n, m, \tilde n, \tilde m|\mathfrak Q, \bar{\mathfrak Q'}, \eta,  j,  n, m, \tilde n, \tilde m) = 1
\eeq
which provides $\mathcal N(r, \rho) =  2e^{2(r^{2} + \rho^{2})}$.
\subsubsection{Resolution of the identity}
\bpro\label{quatresolv000}
The QVCSs (\ref{nqvcs00}) fulfill a resolution of the identity property on $\mathbb C^{2} \otimes \mathcal  H_{q} \otimes \mathcal  H_{q}$ given by
\beq{\label{qvcsresolu}}
&&\sum_{j=1}^{2}\sum_{\tilde m = 0}^{\infty}\sum_{\tilde n = 0}^{\infty}\frac{1}{\tilde m! \tilde n!}  \int_{D_{1} \times D_{2}}
d\mu(\mathfrak Q, \bar{\mathfrak Q'})
(\overrightarrow{\partial_{r}})^{\tilde m }
(\overrightarrow{\partial_{\rho}})^{m}
[\mathcal N(r, \rho)\cr
&&|\mathfrak Q, \bar{\mathfrak Q'}, \eta,  j,  n, m, \tilde n, \tilde m)
(\mathfrak Q, \bar{\mathfrak Q'}, \eta,  j,  n, m, \tilde n, \tilde m|]
(\overleftarrow{\partial_{r}})^{\tilde m}(\overleftarrow{\partial_{\rho}})^{m}
= \mathbb I_{2} \otimes \mathbb I_{q} \otimes \mathbb I_{q}
\eeq
where $d\mu(\mathfrak Q, \mathfrak Q') = \frac{1}{16 \pi^{2}}rdr \rho d\rho (\sin{\phi})d\phi  d\eta d\vartheta (\sin{\varphi})d\varphi
 d\varrho d\gamma$ on $D_{1} \times D_{2};$
\epro
$D_{1} = \{(r,\phi, \eta, \vartheta)| 0 \leq r < \infty,  0 \leq \phi \leq \pi,  0 \leq \eta, \vartheta < 2\pi\}$ and
$D_{2} = \{(\rho,\varphi, \varrho, \gamma)| 0 \leq \rho < \infty,  0 \leq \varphi \leq \pi,  0 \leq \varrho, \gamma < 2\pi\}$.

 The moment problem issued from (\ref{qvcsresolu}), stated as follows:
\beq
\int_{0}^{\infty} \int_{0}^{\infty} \frac{ 4 \pi^{2} W(r, \rho)}{ \mathcal N(r, \rho)}
\frac{r^{2 n}}{ n !}
\frac{\rho^{2 m}}{ m !} rdr  \rho d\rho =  1, \quad W(r, \rho) = \frac{1}{\pi^{2}} \mathcal N(r, \rho)e^{-(r^{2}+\rho^{2})},
\eeq
and is solved.

{\bf Proof.} Similar to the proof of Proposition 4.2 in \cite{a-hk}.

$\hfill{\square}$
\subsection{Dispersions of operators in the QVCSs and uncertainty relations}
This part of the work deals with the physical features of the QVCSs. The  expectation, which can be interpreted as the average
of the observable
 that  would be  expected to obtain from a large number of
measurements, and the dispersion
of the quadrature operators in the constructed QVCSs, are investigated  on the Hilbert space $\mathbb C^{2} \otimes \mathcal  H_{q} \otimes \mathcal  H_{q}$.

Let us consider, by using the Eqs.  (\ref{opchi03}),
the  operators given  on $\mathbb C^{2} \otimes \mathcal  H_{q} \otimes \mathcal  H_{q}$ by
\beq\label{quadra000}
\hat P_{X} &=& \mathbb I_{2} \otimes \frac{-i \hbar }{\sqrt{2 \Theta}}[\mathfrak A_{R} - \mathfrak A^{\dag}_{R}, \ .],
\qquad
\hat P_{Y} =  \mathbb I_{2} \otimes \frac{-\hbar }{\sqrt{2 \Theta}}[\mathfrak A_{R} + \mathfrak A^{\dag}_{R}, \ .],
\cr
\hat X  &=&  \mathbb I_{2} \otimes \sqrt{\frac{\Theta}{2}}[\mathfrak A_{R}+\mathfrak A^{\dag}_{R}],  \quad
\hat Y = \mathbb I_{2} \otimes i \sqrt{\frac{\Theta}{2}}[\mathfrak A^{\dag}_{R} - \mathfrak A_{R}],\quad \Theta = \frac{1}{eB(1-eB\theta)},
\eeq
where $\mathfrak A_{R} \equiv \mathfrak{A}$  and $\mathfrak A_{R}^{\dag} \equiv \mathfrak{A}^{\ddag}$ by acting in the right of a given state $|\tilde n \rangle   \langle \tilde m| \otimes |m\rangle \langle n|$.

From (\ref{hfock000}) and (\ref{opchi03}) together, we obtain
\beq\label{algeb003}
[\mathfrak A^{\dag}_{R} - \mathfrak A_{R}, \ |\tilde n \rangle   \langle \tilde m| \otimes |m\rangle \langle n|] =
\sqrt{n+1}|\tilde n \rangle   \langle \tilde m| \otimes |m\rangle \langle n+1|
\sqrt{n}|\tilde n \rangle   \langle \tilde m| \otimes |m\rangle \langle n|.
\eeq
Then, we have the following result.
\bpro
Defining  the expectation value
of a given  operator acting on $|\tilde n \rangle   \langle \tilde m| \otimes |m\rangle \langle n|$  by
$\displaystyle \langle \cdot \rangle \deq  \sum_{\tilde n, \tilde m  = 0}^{\infty}
(\mathfrak Q, \bar{\mathfrak Q'}, \eta,  j,  n, m, \tilde n, \tilde m|\cdot|\mathfrak Q, \bar{\mathfrak Q'}, \eta,  j,  n, m, \tilde n, \tilde m)$. Then, we  get
the following quantities:
\beq{\label{quad000}}
\langle \hat P_{X} \rangle
 = \frac{\hbar }{\sqrt{2 \Theta}}r\cos{(\phi) \sin{(\vartheta)}},  \qquad \langle \hat P^{2}_{X} \rangle =
\frac{\hbar^{2} }{\Theta}[r^{2}\sin^{2}(\vartheta) +\frac{1}{4}],
\cr
\langle \hat P_{Y} \rangle =  -
\frac{\hbar }{\sqrt{2\Theta}}[r \cos (\vartheta)],  \qquad \langle \hat P^{2}_{Y} \rangle =
\frac{\hbar^{2} }{\Theta}[r^{2}\cos^{2}(\vartheta) + \frac{1}{4}],
\eeq
from which result the dispersions
\beq{\label{quad003}}
(\Delta \hat P_{X} )^{2}
= \frac{\hbar^{2} }{4 \Theta} [4r^{2}\sin^{2}(\vartheta) - 2r^{2}\cos^{2}(\phi)\sin^{2}(\vartheta)  + 1], \
(\Delta \hat P_{Y} )^{2}
= \frac{\hbar^{2}}{2 \Theta} [r^{2}\cos^{2}(\vartheta) + \frac{1}{2}].
\eeq
Thereby,  one gets the following  uncertainties modified
\beq\label{uncert000}
[\Delta \hat X \Delta \hat Y]^{2} &=&  \frac{1}{16}\left(\frac{\Theta^{2}}{4}\right)   F(r,\vartheta,\phi) =
\frac{1}{16} \left[\frac{1}{4}|\langle [\hat X, \hat Y] \rangle|^{2} \right] F(r,\vartheta,\phi),  \cr
[\Delta \hat X  \Delta \hat P_{X}]^{2}  &=&    \frac{1}{16} \left(\frac{\hbar^{2} }{4}\right)
 F(r,\vartheta,\phi) \geq  \frac{1}{16} \left[  \frac{1}{4}|\langle [\hat X, \hat P_{X}]  \rangle|^{2}\right],  \cr
[\Delta \hat Y \Delta \hat P_{Y}]^{2}  & = &   \frac{1}{16} \left(\frac{\hbar^{2} }{4}\right)
 F(r,\vartheta,\phi) \geq  \frac{1}{16} \left[  \frac{1}{4}|\langle [\hat Y, \hat P_{Y}]  \rangle|^{2}\right], \cr
\lim_{\theta \longrightarrow \frac{1}{eB}} [\Delta \hat P_X \Delta \hat P_Y]^{2} &=&
\lim_{\theta \longrightarrow \frac{1}{eB}}
\frac{1}{16} \left[\frac{1}{4}|\langle [\hat P_X, \hat P_Y] \rangle|^{2} \right] F(r, \vartheta,\phi) =  0,
\eeq
where
\beq\label{func007}
F(r,\vartheta,\phi) = [2r^{2}\cos^{2}(\vartheta) +1]
[4r^{2}\sin^{2}(\vartheta) - 2r^{2}\cos^{2}(\phi)\sin^{2}(\vartheta)  +1],
\eeq
\beq
|\langle [\hat X, \hat Y] \rangle|^{2} = \Theta^2, \quad |\langle [\hat X, \hat P_{X}]  \rangle|^{2} = \hbar^{2} = |\langle [\hat Y, \hat P_{Y}]  \rangle|^{2}, \quad
|\langle [\hat P_X, \hat P_Y] \rangle|^{2} = \frac{\hbar^2}{4\Theta^2}.
\eeq
 \epro
  \begin{center}
	\begin{figure}[h]
		\begin{subfigure}[b]{0.3\textwidth}
			\includegraphics[width=\textwidth]{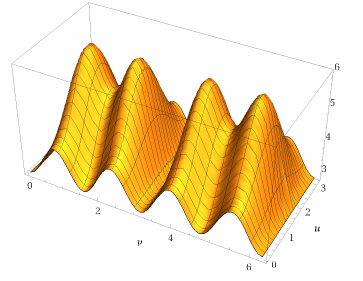}
			\caption{$m=2$}
		\end{subfigure}
	\begin{subfigure}[b]{0.3\textwidth}
		\includegraphics[width=\textwidth]{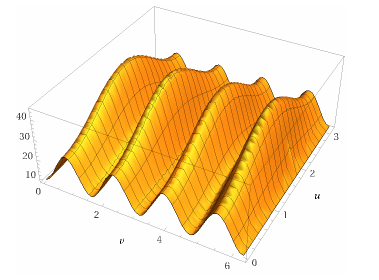}
		\caption{$m=5$}
	\end{subfigure}
\begin{subfigure}[b]{0.3\textwidth}
	\includegraphics[width=\textwidth]{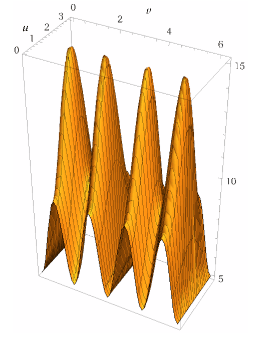}
	\caption{$m=7$}
\end{subfigure}	\caption{\it \small
		Plots of the function $F(r,\vartheta,\phi)$ (\ref{func007}) depending on $r, \vartheta \equiv v \in [0, 2\pi)$, and  $\phi \equiv u \in  [0, \pi]$: (a): $m = 2, r = \sqrt{2}$;  (b): $m = 5, r = \sqrt{2}$;  (c): $m = 7, r = \sqrt{2}$.
		}
	\end{figure}
\end{center}
 there is a double periodicity highlighted by  oscillations that occur along both $u$ and $v$ axes, representing different
combinations of angular parameters.
 The positive definiteness of $F$, i.e., $F \geq 1$ everywhere, ensures physically meaningful uncertainties bounded below by the vacuum limit.
These plots characterize the semiclassical behavior and the effects induced by noncommutativity. The results show that the quadrature dispersions depend explicitly on the continuous parameters $r$, $\vartheta$, and $\phi$, which control the amplitude, phase, and internal orientation of the QVCSs.
The uncertainty products $\Delta X,\Delta Y$, $\Delta X,\Delta P_X$, and $\Delta Y,\Delta P_Y$ satisfy generalized uncertainty relations modified by the noncommutative parameter $\theta$.
 As $\theta \to 1/(eB)$, $\Theta \to \infty$ and $\Delta P_X \Delta P_Y \to 0$, indicating a singular dynamical regime with vanishing effective mass, i.e.,  $M^{*} = M(1-eB\theta)\rightarrow  0$. For $r \to 0$, the function approaches $F(0,\vartheta,\phi) = 1$, representing minimum Heisenberg uncertainty without quantum squeezing. Then, one can compare
  the elevated floor $F_{\min} > 1$ as indicating  that one quadrature is persistently squeezed while the conjugate quadrature is anti-squeezed, preserving the Heisenberg bound \cite{gerry}.
\section{Time evolution   and density of probability}
The QVCSs $|\mathfrak Q, \bar{\mathfrak Q'}, \eta,  j,  n, m, \tilde n, \tilde m)$, given the shifted  Hamiltonian $\mathbb H = \hat H  -  \frac{\hbar \omega^{*}}{2}\mathbb I_{\mathcal F_{\mathcal K}}$  with spectrum $\mathcal E_{n} =  \omega^{*} n,   \; \hbar = 1$, see (\ref{eigval00}),
satisfy, under the time evolution operator
$\mathbb U(t) =  e^{-i  t \mathbb H}$,  the following  property:
\beq\label{timev00}
\mathbb U(t)|\mathfrak Q, \bar{\mathfrak Q'}, \eta,  j,  n, m, \tilde n, \tilde m) =
|\mathfrak Qe^{-i  t     \omega^{*}}, \bar{\mathfrak Q'}, \eta,  j,  n, m, \tilde n, \tilde m) = |\mathfrak Q(t), \bar{\mathfrak Q'}, \eta,  j,  n, m, \tilde n, \tilde m),
\eeq
where $\mathfrak Q = A(r)e^{i \vartheta \Theta(\hat n )} = r\mathbb I_2e^{i \vartheta \Theta(\hat n )}$ with  $\mathfrak Q(t) \deq   {\mathfrak Q}e^{-i \omega^{*} t}
= r\mathbb I_2e^{i (\vartheta \Theta(\hat n ) - \omega^{*}t\mathbb I_2)}$ such that $ \mathfrak Q(t) =  r\left[\cos{(\vartheta - \omega^{*}t)}\mathbb I_2 + i\Theta(\hat n)\sin{(\vartheta - \omega^{*}t)}
\right]$ providing
{\small
{\small\beq\label{quaterntransf}
{\mathfrak Q }e^{-i  t     \omega^{*}}
= \left(\begin{array}{cc}
 r[\cos (\vartheta - \omega^{*}t) + i \cos (\phi) \sin (\vartheta - \omega^{*}t)] &
i r e^{i \eta} \sin (\phi) \sin (\vartheta - \omega^{*}t) \\
i  r  e^{-i \eta} \sin (\phi) \sin (\vartheta - \omega^{*}t) &
 r[\cos (\vartheta - \omega^{*}t) - i \cos (\phi) \sin (\vartheta - \omega^{*}t)] \\
                \end{array}
\right).
\eeq}
}
Using the result of the QVCSs overlap $ (\mathfrak Q, \bar{\mathfrak Q'}, \eta,  j,  n, m, \tilde n, \tilde m|\mathfrak Q_0, \bar{\mathfrak Q'}, \eta,  j,  n, m, \tilde n, \tilde m)$,
 since $\mathfrak Q'$ commutes with $\mathfrak Q_0$ and $\mathfrak Q$ together, we get
{\small\beq
&&\left|\sum_{j=1}^{2}\sum_{\tilde n, \tilde m = 0}^{\infty}(\mathfrak Q, \bar{\mathfrak Q'}, \eta,  j,  n, m, \tilde n, \tilde m|\mathfrak Q_0, \bar{\mathfrak Q'}, \eta,  j,  n, m, \tilde n, \tilde m) \right|^2\cr
&=& [\mathcal N(r,\rho)]^{-1}[\mathcal N(r_0, \rho)]^{-1}\sum_{\tilde  m, \tilde k = 0}^{\infty}\frac{Tr(|\mathfrak Q'|^{4m})  Tr(|\mathfrak Q'|^{2\tilde m}) Tr(|\mathfrak Q'|^{2\tilde k})}{R^2(m)R(\tilde  m)R(\tilde  k)}\cr
&&\times \sum_{k = 0}^{\infty}\sum_{n = 0}^{\infty}\frac{Tr(|\mathfrak Q|^{2k}|\mathfrak Q_0|^{2n}) }{R(k)R(n)}Tr\left(e^{(\bar{\mathfrak Q}_0\mathfrak Q) + (\mathfrak Q_0\bar{\mathfrak Q})}\right).
\eeq}
Then, with $\mathfrak Q = A(r)e^{i \vartheta \Theta(\hat n )} = r\mathbb I_2e^{i \vartheta \Theta(\hat n )}$ and $\mathfrak Q_0 = B(r_0)e^{i \vartheta_0 \tilde \Theta(\hat k )} = r_0\mathbb I_2e^{i \vartheta_0 \tilde \Theta(\hat k )}$, we obtain
\beq
Tr\big(e^{\bar{\mathfrak Q}_0\mathfrak Q + \mathfrak Q_0\bar{\mathfrak Q}}\big) = 4e^{2 r_0 r \cos \vartheta_0  \cos \vartheta} \cos\big( 2 r_0 r \sin \vartheta_0 \sin \vartheta \big).
\eeq
Thereby, the time evolution behavior of $\varrho_{\mathfrak Q_0 }(\mathfrak Q, t)$ is provided by
{\small\beq\label{density000}\label{qvcstempdens000}
\mathfrak Q \mapsto \varrho_{\mathfrak Q_0 }(\mathfrak Q,t) &\deq&
\left|\sum_{j=1}^{2}
\sum_{\tilde n, \tilde m = 0}^{\infty}(\mathfrak Q, \bar{\mathfrak Q'}, \eta,  j,  n, m, \tilde n, \tilde m|\mathbb U(t)|\mathfrak Q_0, \bar{\mathfrak Q'}, \eta,  j,  n, m, \tilde n, \tilde m) \right|^2 \cr
&=& \frac{2}{\sqrt{{\mathcal N}(\rho, \rho)}}\left(\frac{\rho^{2m}}{m !}\right)^2
 \frac{\left\{4e^{2 r_0 r \cos  (\vartheta_0 - \omega^{*}t)  \cos \vartheta} \cos\left( 2 r_0 r \sin (\vartheta_0 - \omega^{*}t) \sin \vartheta\right)\right\}}{\sqrt{\mathcal N(r, r_0)}}\nonumber\\
\eeq}
with
\beq
\mathfrak Q_0(t)
=   r_0(t)\left[\cos{(\vartheta_0 - \omega^{*}t)}\mathbb I_2 + i\Theta_0(\hat n)\sin{(\vartheta_0 - \omega^{*}t)}
\right] \deq  \mathfrak Q_0 e^{-i  t    \omega^{*}},
\eeq
where $\vartheta_0,  \Theta_0(\hat n )$ are
given as in (\ref{quat007}).

\begin{figure}[h]
	\begin{subfigure}[b]{0.3\textwidth}
		\includegraphics[width=\textwidth]{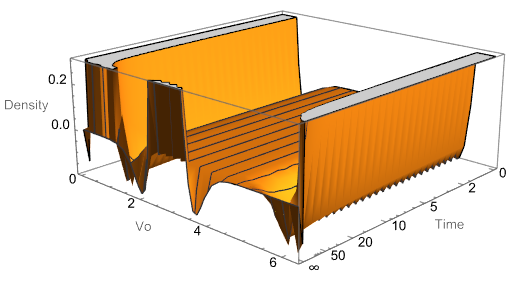}
		\caption{$V=\vartheta$}
	\end{subfigure}
	\begin{subfigure}[b]{0.3\textwidth}
		\includegraphics[width=\textwidth]{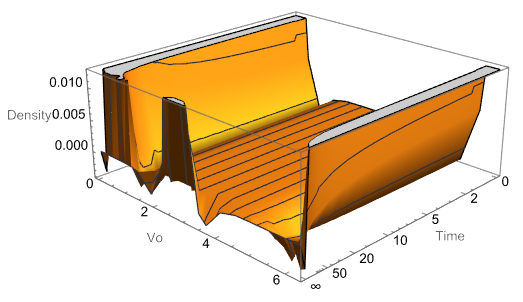}
		\caption{$V=\vartheta$}
	\end{subfigure}
	\begin{subfigure}[b]{0.3\textwidth}
		\includegraphics[width=\textwidth]{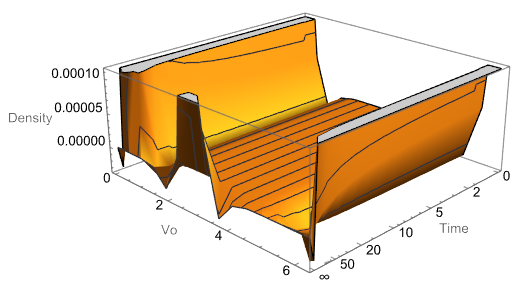}
		\caption{$V=\vartheta$}
	\end{subfigure}
	\begin{subfigure}[b]{0.3\textwidth}
		\includegraphics[width=\textwidth]{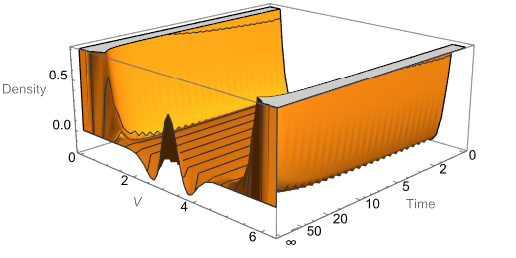}
		\caption{$V_0=\vartheta_0$}
	\end{subfigure}
	\begin{subfigure}[b]{0.3\textwidth}
		\includegraphics[width=\textwidth]{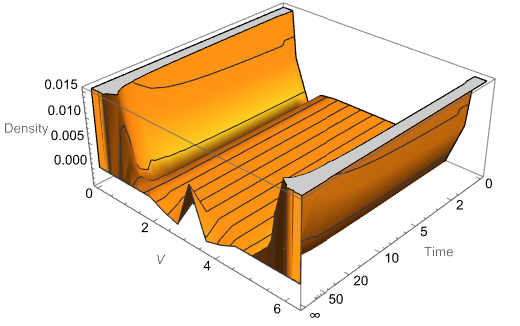}
		\caption{$V_0=\vartheta_0$}
	\end{subfigure}
	\begin{subfigure}[b]{0.3\textwidth}
		\includegraphics[width=\textwidth]{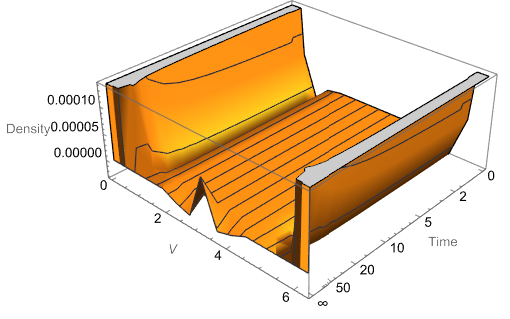}
		\caption{$V_0=\vartheta_0$}
	\end{subfigure}
		\caption{\it \small Plots of the  temporal probability density $\varrho_{\mathfrak{Q}_0}(\mathfrak{Q},t)$ (\ref{qvcstempdens000}) versus $V_0(t) \equiv \vartheta_0(t) = \vartheta_0 - \omega^{*}t \in [0, 2\pi), t \in [0, \infty)$ (in normalized units) and $V\equiv\vartheta \in [0, 2\pi)$,  with $ \omega^{*} = 2.5. 10^{-3}$(in normalized units) and $r_0, r$ and $\rho$ fixed:(a): $m = 2, \vartheta =\pi/6 $; (b): $m = 5, \vartheta =\pi/6 $; (c): $m = 7, \vartheta =\pi/6 $; (d): $m = 2, \vartheta_0= \pi/6$; (e): $m = 5, \vartheta_0 =\pi/6 $; (f): $m = 7, \vartheta_0 = \pi/6$.
	.
		}
	\end{figure}

Figure 4 displays the temporal probability density of the QVCSs. The first three graphs (a), (b) and (c) display $\varrho_{\mathfrak Q_0 }(\mathfrak Q,t) $ for quantum numbers $m = 2, 5, 7$ respectively, with the spatial angle $\vartheta$ held constant. For $m = 2$, the probability density  exhibits a regular periodic modulation in  $\vartheta_0 $, with well-defined maxima and relatively large amplitude, $\rho_{\max} \approx 0.2$,  reflecting strong coherence and predominantly constructive interference, with sharp periodic peaks appear at specific values of $\vartheta_0$,  in a weakly excited, quasi-classical regime.
 At $m = 5$, the overall amplitude decreases, with $\rho_{\max} \approx 0.01$,  due to the factor $(\rho^{2m}/m!)^2$, while
 for $m = 7$, the probability density is strongly suppressed, followed by a reduction to $\rho_{\max} \approx 10^{-4}$ indicating significant wave packet
spreading due to increased quantum number $m$.
By fixing the initial angle $\vartheta_0$ while exploring the spatial
angle $\vartheta$, we obtain for $m=2$, a maximum density $\rho_{\max} \approx 0.5$ which appears as sharp ridges in the $(\vartheta, t)$ plane,  indicating strong spatial localization at specific angles. For $m=5$, we have a reduced density scale ($\rho_{\max} \approx 0.015$), while for $m=7$, we get an ultra-low density ($\rho_{\max} \approx 10^{-4}$). For $t \to \infty$, the density spreads across all angular values, approaching a quasi-uniform distribution characteristic of long-time decoherence in open quantum systems.  The peak positions in $\varrho_{\mathfrak Q_0 }(\mathfrak Q,t) $ can be associated to transitions between Landau levels modified by noncommutativity, while experimental  realization through scanning tunneling spectroscopy on graphene or topological insulators could probe these structures  \cite{girvin}.
\brmk
Note that it might be of interest to carry out the following procedure on a separable
abstract left or right quaternionic  Hilbert space as developed in, for example, \cite{mural-santhar}.
\ermk
\section{Multimatrix vector coherent states  from unitary maps}
In this section, we develop an alternative construction of quaternionic vector coherent states
(QVCSs) by employing unitary transformations associated with the Wigner representation.
This approach enables us to realize the QVCSs on an extended Hilbert space structure and
establish their completeness properties through an explicit resolution of the identity.

Consider the unitary map
$U(x,y)$ on $\mathcal B_{2}(\mathfrak H)$ given by
\begin{eqnarray}\label{unitp001}
(U(x,y)\Phi)(\xi) =  e^{-i x \left(\xi - y/2 \right)}\Phi\left(\xi -
y \right),
\end{eqnarray}
with  $U(x,y) = e^{-i (xQ + y P)}$,   $Q$ and $P$ being the usual position and momentum operators in the Schr\"{o}dinger
representation satisfying $[Q, P] = i \mathbb I_{\mathfrak H_{s}}$, where $\mathfrak H_{s} = L^{2}(\R^{2}, dxdy)$.
Given any vector $X \in \mathcal B_{2}(\mathfrak H)$,
$X = |\Phi \rangle \langle \Psi|$, one has
\begin{eqnarray}{\label{map1}}
 && \mathcal W: \mathcal B_{2}(\mathfrak H) \rightarrow L^{2}(\R^{2}, dxdy) \cr
 \cr
(\mathcal WX)(x,y) &\deq& \frac{1}{(2\pi)^{1/2}} Tr\left[U(x,y)^{*}X\right] = \frac{1}{(2\pi)^{1/2}} \langle U(x,y)\Psi|\Phi \rangle_{\mathfrak H} \cr
&=& \frac{1}{(2\pi)^{1/2}}\int_{\R}e^{i x\left(\xi -
y/2 \right)}\overline{\Psi\left(\xi- y \right)}\Phi(\xi)d\xi.
\end{eqnarray}
The mapping $\mathcal W$ is often referred to as the Wigner transform in the physical literature and it is well known to be unitary \cite{ali-antoine-gazeau}. The inverse of $\mathcal W$ \cite{iaremuaetal} is defined on the dense set of vectors $f \in L^{2}(\R^{2}, dxdy)$, comprising the image of $\mathfrak H \otimes \overline{\mathfrak H} \simeq \mathcal B_{2}(\mathfrak H)$, the inverse map $\mathcal W^{-1}$ is such that
\begin{eqnarray}
&&\mathcal W^{-1}:L^{2}(\R^{2}, dxdy) \rightarrow \mathfrak H \otimes \overline{\mathfrak H} \cr
&&\mathcal W^{-1} f =  \int_{\R}\int_{\R}U(x,y)\mathcal W(|\phi\rangle \langle \psi|)(x,y)dxdy.
\end{eqnarray}
Consider the composite  map
\begin{eqnarray}\label{unitp003}
\mathcal U:L^{2}(\R^{2}, dxdy) \rightarrow \mathfrak H \otimes \mathfrak H = L^{2}(\R) \otimes L^{2}(\R),
\end{eqnarray}
with $\mathcal U = \mathcal I \circ \mathcal W^{-1}$, where $\mathcal I:\mathfrak H \otimes \overline{\mathfrak H} \rightarrow \mathfrak H \otimes \mathfrak H$, such that for a given vector $|\phi\rangle \langle \psi| \in \mathfrak H \otimes \overline{\mathfrak H}$, $\mathcal I(\phi(x)\overline{\psi(y)}) = \phi(x)\psi(y), \; x,y \in \R, \phi, \psi \in \mathfrak H$. Next, introduce the  antiunitary operator defined in \cite{aremua-gouba} as
\beq\label{unitp005}
\mathcal J: \mathcal B_{2}(\mathfrak H) \longrightarrow \mathcal B_{2}(\mathfrak H), \quad \mathcal J (|\phi\rangle\langle \psi|) = |\psi\rangle\langle \phi|, \quad  \forall  |\phi\rangle, |\psi\rangle \in \mathfrak H,
\eeq
and then let $\tilde{\mathcal U} = \mathcal J \circ \mathcal W^{-1}$.
In order to obtain mapped states via unitary transformations to larger Hilbert spaces, let  us define the unitary transformations operators from
$\{|n, m, \tilde n, \tilde m), n, m, \tilde n, \tilde m \in \N \}$ to $\{|\Psi_{n, m, \tilde n, \tilde m}), n, m, \tilde n, \tilde m\in \N \}$, with $|n, m, \tilde n, \tilde m):=|\tilde n \rangle   \langle \tilde m| \otimes |m\rangle \langle n| $,  and vice versa given by
\beq{\label{spec00}}
\mathcal V|n, m, \tilde n, \tilde m) = |\Psi_{n, m, \tilde n, \tilde m}), \qquad \tilde{\mathcal V}|\Psi_{n, m, \tilde n, \tilde m}) = |n, m, \tilde n, \tilde m)
\eeq
where their expansions write
\beq{\label{spec01}}
\mathcal V = \sum_{n, m, \tilde n, \tilde m=0}^{\infty} |\Psi_{n, m, \tilde n, \tilde m})(n, m, \tilde n, \tilde m|, \qquad \tilde{\mathcal V} = \sum_{n, m, \tilde n, \tilde m=0}^{\infty} |n, m, \tilde n, \tilde m)(\Psi_{n, m, \tilde n, \tilde m}|,
\eeq
respectively. One gets
\beq
\mathcal V \tilde{\mathcal V} &=& \sum_{n, m, \tilde n, \tilde m=0}^{\infty}|\Psi_{n, m, \tilde n, \tilde m})(\Psi_{n, m, \tilde n, \tilde m}| = \1_{q} \otimes \1_{q}, \cr
\, \tilde{\mathcal V} \mathcal V &=& \sum_{n, m, \tilde n, \tilde m=0}^{\infty}|n, m, \tilde n, \tilde m)(n, m, \tilde n, \tilde m| =  \1_{q} \otimes \1_{q}.
\eeq
Let the two classes of QVCSs
\beq\label{unitopqvcs000}
\tilde{\mathcal U}[\mathcal W\left\{\mathcal V|\mathfrak Q, \mathfrak Q',  \eta, j, n, m, \tilde n, \tilde m)\right\}] (x',y', x,y) &\deq& |\xi_{\mathfrak Q, \mathfrak Q'} (\Phi,\Psi)\rangle, \crcr \mathcal U[\mathcal W\left\{\mathcal V|\mathfrak Q, \mathfrak Q',  \eta, j, n, m, \tilde n, \tilde m)\right\}] (x',y', x,y) &\deq& |\eta_{\mathfrak Q, \mathfrak Q'} (\Phi,\Psi)\rangle,
\eeq
constructed from the QVCSs $|\mathfrak Q, \mathfrak Q',  \eta, j, n, m, \tilde n, \tilde m)$, defined on $\mathbb{C}^2 \otimes \mathcal{H}_q \otimes \mathcal{H}_q$, on  the Hilbert space  $\C^2 \otimes \mathfrak{H}^{\otimes 4}$  with $\mathfrak{H} = L^2(\R)$, given by
\beq
 |\xi_{\mathfrak Q, \mathfrak Q'} (\Phi,\Psi)\rangle
 &=& (\mathcal N(r, \rho))^{-1/2} \sum_{n,m=0}^{\infty}F_{n}(\mathfrak Q)F_{m}(\mathfrak Q')e^{-i \eta \mathcal E_{n}} \chi^{j}\otimes \Psi_{\tilde m}(x')
\otimes \overline{\Phi}_{\tilde n}(y') \cr
&& \otimes \Psi_{n}(x)
\otimes \overline{\Phi}_{m}(y),
\cr
|\eta_{\mathfrak Q, \mathfrak Q'} (\Phi,\Psi)\rangle
&=& (\mathcal N(r, \rho))^{-1/2} \sum_{n,m=0}^{\infty}F_{n}(\mathfrak Q)F_{m}(\mathfrak Q')e^{-i \eta \mathcal E_{n}} \chi^{j} \otimes \Psi_{\tilde m}(x')
\otimes \overline{\Phi}_{\tilde n}(y')\cr
&&\otimes \Psi_{n}(x)
\otimes \overline{\Phi}_{m}(y).
\eeq
We have the following result:
\bpro\label{unitprop000}
The QVCSs $|\xi_{\mathfrak Q, \mathfrak Q'} (\Phi,\Psi)\rangle$ and $|\eta_{\mathfrak Q, \mathfrak Q'} (\Phi,\Psi)\rangle$, provided the completeness
 relations on $\{|x\rangle\}$ and $\{|y\rangle\}$ representations given by $\displaystyle \int_{\R}| x\rangle \langle x|dx = I_{\mathfrak H} = \int_{\R}| y\rangle \langle y|dy$ with $\mathfrak H = L^2(\R)$,  satisfy on  $\C^2 \otimes \mathfrak{H}^{\otimes 4}$ the  resolutions of the identity
\beq
&&\sum_{j=1}^{2}\sum_{m, \tilde m, \tilde n, n=0}^{\infty}\int_{D_1\times D_2}\int_{D_1\times D_2}\int_{\R}\int_{\R}\int_{\R}\int_{\R}\mathcal N(r, \rho)(d\mu(\mathfrak Q, \mathfrak Q'))^2|\xi_{\mathfrak Q, \mathfrak Q'} (\Phi,\Psi)\rangle\langle\xi_{\mathfrak Q, \mathfrak Q'} (\Phi,\Psi)|\cr
&& \times dxdydx'dy'= \mathbb I_{2}\otimes I_{\mathfrak{H}^{\otimes 4}},
\cr
\cr
&&\sum_{j=1}^{2}\sum_{n, m, \tilde n, \tilde m=0}^{\infty}\int_{D_1\times D_2}\int_{D_1\times D_2}\int_{\R}\int_{\R}\int_{\R}\int_{\R}\mathcal N(r, \rho)(d\mu(\mathfrak Q, \mathfrak Q'))^2|\eta_{\mathfrak Q, \mathfrak Q'} (\Phi,\Psi)\rangle\langle\eta_{\mathfrak Q, \mathfrak Q'} (\Phi,\Psi)|\cr
&&\times dxdydx'dy' = \mathbb I_{2}\otimes I_{\mathfrak{H}^{\otimes 4}},
\eeq
respectively, where
$\mathcal D = D_1 \times D_2$ with
$D_{1} = \{(r,\phi, \eta, \vartheta)| 0 \leq r < \infty,  0 \leq \phi \leq \pi,  0 \leq \eta, \vartheta < 2\pi\}$ and
$D_{2} = \{(\rho,\varphi, \varrho, \gamma)| 0 \leq \rho < \infty,  0 \leq \varphi \leq \pi,  0 \leq \varrho, \gamma < 2\pi\}$.
\epro

{\bf Proof.} See in the Appendix.

$\hfill{\square}$
\section{Concluding remarks}\label{sec5}
	In this work, we have developed a theory of coherent states for the exotic Landau problem, an advanced quantum mechanical system characterized by quantum states in noncommutative space enriched with internal degrees of freedom. Our approach begins with the derivation of the classical counterpart, formulated through modified Poisson brackets and characterized by conserved quantities $\mathcal{P}_i$ and $\mathcal{K}_i$ with $i = 1,2$. The quantum Fock space naturally emerges as the tensor product of two chiral oscillator sectors, $\mathcal{F}_{\mathcal{K}}$ and $\mathcal{F}_{\mathcal{P}}$, each sector being intimately connected to the underlying conserved quantities.

	Building upon this foundation, we have constructed coherent states on the quantum Hilbert space $\mathcal{H}_q$ that rigorously satisfy all Klauder criteria. These coherent states provide an interesting  mathematical framework that unifies quantum optics, noncommutative geometry, and quantum information theory, revealing several remarkable quantum behaviors. We have calculated  the free particle propagator that manifests UV regularization as a direct consequence of spatial noncommutativity, and we have  thoroughly analyzed the nonclassical characteristics of time evolution and photon number distribution.

	A significant extension of our work involves the construction of vector and quaternionic vector coherent states. These are realized by introducing vector states labeled by multiple quantum numbers $(n, m, \tilde{n}, \tilde{m})$ together with an internal index $j = 1, 2$, thereby capturing a sophisticated quantum system possessing both spatial and internal structure \cite{perelomov}. We have derived  uncertainty relations specific to the quaternionic vector coherent state representation,  and analyzed  their dynamical evolution. Through the Wigner transform, we have established unitary mappings from the quaternionic vector coherent states to expanded Hilbert spaces, particularly to $\mathbb{C}^2 \otimes \mathfrak{H}^{\otimes 4}$ with $\mathfrak{H} = L^2(\mathbb{R})$.

	The theoretical framework presented here opens avenues for investigating quantum information protocols and noncommutative field theories. Previous work has already applied the exotic Landau problem to qubit teleportation \cite{aremua-gouba-qubit}. Our coherent state construction holds significant potential for quantum information applications: varying the parameter $m$ yields different encoding strategies, each exhibiting distinct robustness against environmental decoherence \cite{terhal}. Furthermore, the $m$-dependent phenomenology uncovered in our study points toward excitation-number-driven quantum phase transitions \cite{sachdev}.
%

\section*{Appendix}
{\bf Proof of Proposition (\ref{contin000})}

 From the CSs (\ref{cohst00}) definition,   the term $|||z,\bar{z}';m) - |z',\bar{z}'',m)||^{2}_{\mathcal{HS}}$ in the proposition is evaluated as
\beq\label{qtrace000}
|||z,\bar{z}';m) - |z',\bar{z}'',m)||^{2}_{\mathcal{HS}}
&=&
|||z,\bar{z}';m)||^{2}_{\mathcal HS} + |||z',\bar{z}'',m)||^{2}_{\mathcal HS} - (z,\bar{z}';m|z',\bar{z}'',m)\cr
&&- (z',\bar{z}'',m|z,\bar{z}';m) \cr
&=& 2  - tr_{c}[(|z\rangle \langle z| \otimes |\bar{z}'\rangle \langle \bar{z}'|)^{\dag}(|z'\rangle \langle z'|\otimes|\bar{z}''\rangle \langle \bar{z}''|)] \cr
&&- tr_{c}[(|z'\rangle \langle z'|\otimes|\bar{z}''\rangle \langle \bar{z}''|)^{\dag}(|z\rangle \langle z|\otimes|\bar{z}'\rangle \langle \bar{z}'|)]
\eeq
where we set
\beq
\mathfrak{Trc}_1 &=&
tr_{c}[(|z\rangle \langle z|\otimes|\bar{z}'\rangle \langle \bar{z}'|)^{\dag}(|z'\rangle \langle z'|\otimes|\bar{z}''\rangle \langle \bar{z}''|)]\cr
&=& \left[e^{-i z \wedge z'}e^{-\frac{|z-z'|^{2}}{2}}\right]\left[e^{-i z' \wedge z}e^{-\frac{|z'-z|^{2}}{2}}\right]\left[e^{-i \bar{z}' \wedge \bar{z}'' }e^{-\frac{|\bar{z}'-\bar{z}''|^{2}}{2}}\right]\left[e^{-i \bar{z}'' \wedge \bar{z}'}e^{-\frac{|\bar{z}''-\bar{z}'|^{2}}{2}}\right]\cr
 &=& e^{-|z-z'|^{2}}e^{-|\bar{z}'-\bar{z}''|^{2}},
\cr
\mathfrak{Trc}_2 &=&
tr_{c}[(|z'\rangle \langle z'|\otimes|\bar{z}''\rangle \langle \bar{z}''|)^{\dag}(|z\rangle \langle z|\otimes|\bar{z}'\rangle \langle \bar{z}'|)]\cr
 &=& e^{-|z-z'|^{2}}e^{-|\bar{z}'-\bar{z}''|^{2}}.
\eeq
Thereby
\beq
\lim_{z\rightarrow z', z'\rightarrow z''}||z,\bar{z}';m) - |z',\bar{z}'',m)||^{2}_{\mathcal{HS}}
&=& \lim_{z\rightarrow z', z'\rightarrow z''} 2(1-e^{-|z-z'|^{2}}e^{-|\bar{z}'-\bar{z}''|^{2}})\cr
&=& 0
\eeq
if and only if  $|z - z'|$ and $|\bar{z}' - \bar{z}''|$ are sufficiently small.

$\hfill{\square}$

{\bf Proof of Proposition (\ref{unitprop000})}

Using the definition of the QVCSs in (\ref{unitopqvcs000}), we have
\beq
|\xi_{\mathfrak Q, \mathfrak Q'} (\Phi,\Psi)\rangle\langle\xi_{\mathfrak Q, \mathfrak Q'} (\Phi,\Psi)|
&=&\sum_{p, \tilde k=0}^{\infty}\sum_{n, m=0}^{\infty} (\mathcal N(r, \rho))^{-1}F_{p}\bar{(\mathfrak Q)}F_{\tilde k}\bar{(\mathfrak Q')}F_{n}(\mathfrak Q)F_{m}(\mathfrak Q')\cr
&& \times e^{i \eta (\mathcal E_{p} - \mathcal E_{n})}|\chi^{j}\rangle\langle \chi^{j}|\otimes
\langle \Psi_{\tilde k}|x'\rangle \langle x'|\Psi_{\tilde n}\rangle\otimes  \langle \Phi_{\tilde m}|y'\rangle \langle y'|\Phi_{p}\rangle \cr
&& \otimes \langle \Psi_{k}| x\rangle \langle x|\Psi_{n}
\rangle\otimes \langle \Phi_{m}|y\rangle \langle y|\Phi_{\tilde p}\rangle
\eeq
such that
\beq
&&\int_{\R}\int_{\R}\int_{\R}\int_{\R}|\xi_{\mathfrak Q, \mathfrak Q'} (\Phi,\Psi)\rangle\langle\xi_{\mathfrak Q, \mathfrak Q'} (\Phi,\Psi)|dxdydx'dy'\cr
&=& \sum_{p, \tilde k=0}^{\infty}\sum_{n, m=0}^{\infty} (\mathcal N(r, \rho))^{-1}F_{p}\bar{(\mathfrak Q)}F_{\tilde k}\bar{(\mathfrak Q')}F_{n}(\mathfrak Q)F_{m}(\mathfrak Q')\cr
&& \times e^{i \eta (\mathcal E_{p} - \mathcal E_{n})}|\chi^{j}\rangle\langle \chi^{j}| \otimes
\delta_{\tilde k, \tilde n}I_{\mathfrak{H}}\otimes\delta_{\tilde m, p}I_{\mathfrak{H}}\otimes\delta_{k, n}I_{\mathfrak{H}}\otimes\delta_{m, \tilde p}I_{\mathfrak{H}}\cr
&=& (\mathcal N(r, \rho))^{-1}F_{\tilde m}\bar{(\mathfrak Q)}F_{\tilde n}\bar{(\mathfrak Q')}F_{k}(\mathfrak Q)F_{\tilde p}(\mathfrak Q')e^{i \eta (\mathcal E_{\tilde m} - \mathcal E_{k})}|\chi^{j}\rangle\langle \chi^{j}|\otimes I_{\mathfrak{H}}\otimes I_{\mathfrak{H}}\otimes I_{\mathfrak{H}}\otimes I_{\mathfrak{H}}.\quad
\eeq
Thereby
{\small\beq
&&\sum_{j=1}^{2}\sum_{m, \tilde m, \tilde n, n=0}^{\infty}\int_{D_1\times D_2}\int_{D_1\times D_2}\int_{\R}\int_{\R}\int_{\R}\int_{\R}\mathcal N(r, \rho)(d\mu(\mathfrak Q, \mathfrak Q'))^2|\xi_{\mathfrak Q, \mathfrak Q'} (\Phi,\Psi)\rangle\langle\xi_{\mathfrak Q, \mathfrak Q'} (\Phi,\Psi)|\cr
&&\times dxdydx'dy'\cr
&&= \sum_{m, \tilde m, \tilde n, n=0}^{\infty}\frac{1}{16\pi^4}\int_{0}^{2\pi} \int_{0}^{2\pi}\int_{0}^{2\pi}\int_{0}^{2\pi}\int_{0}^{\pi}\int_{0}^{\pi}\left\{(\sin{\phi})(\sin{\varphi}) e^{-i(\tilde m-k)\vartheta \Theta(\hat n)}e^{i(\tilde  n-\tilde p)\gamma \tilde \Theta(\hat k)}\right. \cr
 && \times\left. d\phi  d\eta d\vartheta d\varphi
 d\varrho d\gamma\right\} \frac{1}{16\pi^4}\int_{0}^{2\pi} \int_{0}^{2\pi}\int_{0}^{2\pi}\int_{0}^{2\pi}\int_{0}^{\pi}\int_{0}^{\pi}\left\{(\sin{\phi})d\phi  d\eta d\vartheta (\sin{\varphi})d\varphi
 d\varrho d\gamma\right.\cr
&&\times \left.e^{i(\tilde m-\tilde k)\vartheta \Theta(\hat n)}e^{-i(n-p)\gamma \tilde \Theta(\hat k)}\right\}\int_{0}^{\infty} \int_{0}^{\infty}rdr \rho d\rho\int_{0}^{\infty}\int_{0}^{\infty}rdr \rho d\rho
 \cr
&& \times 
{{ \left(\begin{array}{cc}
\frac{r^{\tilde m} r^{m}}{\sqrt{\tilde m! m!}}\frac{\rho^{\tilde n}\rho^{n}}{\sqrt{\tilde n! n!}} \frac{r^{k} r^{\tilde k}}{\sqrt{k! \tilde k!}}\frac{\rho^{\tilde  p}\rho^{p}}{\sqrt{\tilde p! p!}} \qquad \qquad\
0 \\
0 \qquad \qquad
\frac{r^{\tilde m} r^{m}}{\sqrt{\tilde m! m!}}\frac{\rho^{\tilde n}\rho^{n}}{\sqrt{\tilde n! n!}} \frac{r^{k} r^{\tilde k}}{\sqrt{k! \tilde k!}}\frac{\rho^{\tilde  p}\rho^{p}}{\sqrt{\tilde p! p!}}
                                \end{array}
\right)}}  e^{i \eta (\mathcal E_{\tilde m} - \mathcal E_{k})}\mathbb I_{2}\otimes I_{\mathfrak{H}^{\otimes 4}}
\eeq}
since
\beq
&&\int_{0}^{2\pi}\int_{0}^{2\pi}\int_{0}^{2\pi}\int_{0}^{2\pi}\int_{0}^{\pi}\int_{0}^{\pi} (\sin{\phi})d\phi  d\eta d\vartheta (\sin{\varphi})d\varphi
 d\varrho d\gamma  e^{-i(\tilde m-k)\vartheta \Theta(\hat n)}\cr
 && \times e^{i(\tilde  n-\tilde p)\gamma \tilde \Theta(\hat k)}
 =  \left\{
              \begin{array}{lll}
              0 \quad \mbox{if}  \quad \tilde m \neq k  \, \mbox{and} \,\tilde n \neq \tilde p , \\
               \\
              64\pi^{4} \mathbb I_{2}
\quad \mbox{if}  \quad \tilde m = k  \, \mbox{and} \,\tilde n = \tilde p,
               \end{array}
\right.
\eeq

with the following moment problems
\beq
4 \displaystyle\int_{0}^{\infty} \int_{0}^{\infty}e^{-(r^{2}+\rho^{2})}\frac{r^{2k}}{k!}\frac{\rho^{2\tilde p}}{\tilde p!} rdr \rho d\rho& = 1, \quad 4 \displaystyle\int_{0}^{\infty} \int_{0}^{\infty}e^{-(r^{2}+\rho^{2})}\frac{r^{2\tilde k}}{\tilde k!}\frac{\rho^{2p}}{p!} rdr \rho d\rho = 1
\eeq
satisfied, the proof is completed.

$\hfill{\square}$

\end{document}